\documentclass[useAMS,usenatbib]{mn2e}

\usepackage{amsmath}
\usepackage{amssymb}
\usepackage{soul}
\usepackage{natbib}
\usepackage{graphicx}
\usepackage{caption}
\usepackage{subcaption}
\usepackage{enumerate}
\usepackage{datetime}
\usepackage{color}
\usepackage{hyperref}
\usepackage{draftcopy}

\begin{document}

\title[Radiative Disk Ablation in O \& B Stars]
{Line-driven ablation of circumstellar disks: I. Optically thin decretion disks of classical Oe/Be stars}

 \author[N. D. Kee et al.]
{Nathaniel Dylan Kee$^1$\thanks{Email: dkee@udel.edu},
 Stanley Owocki$^1$, and
 J.O. Sundqvist$^{2,3}$\\
 $^1$ Department of Physics and Astronomy, Bartol Research Institute,
 University of Delaware, Newark, DE 19716, USA\\
 $^2$ Centro de Astrobiologia, 
 Instituto Nacional de Tecnica Aerospacial,
 28850 Torrejon de Ardoz,
 Madrid, Spain\\
 $^3$ Instituut voor Sterrenkunde, KU Leuven, Celestijnenlaan 200D, 3001 Leuven, Belgium
 }

\def\<<{{\ll}}
\def\>>{{\gg}}
\def\wig{{\sim}}
\def\spose#1{\hbox to 0pt{#1\hss}}
\def\+/-{{\pm}}
\def\=={{\equiv}}
\def\mubar{{\bar \mu}}
\def\mustar{\mu_{\ast}}
\def\Lambar{{\bar \Lambda}}
\def\Rstar{r_{\ast}}
\def\Mstar{M_{\ast}}
\def\Lstar{L_{\ast}}
\def\Tstar{T_{\ast}}
\def\gstar{g_{\ast}}
\def\vth{v_{th}}
\def\grad{g_{rad}}
\def\glines{g_{lines}}
\def\Mdot{\dot M}
\def\mdot{\dot m}
\def\yr{{\rm yr}}
\def\ksec{{\rm ksec}}
\def\kms{{\rm km/s}}
\def\qad{\dot q_{ad}}
\def\qlines{\dot q_{lines}}
\def\solar{\odot}
\def\Msun{M_{\solar}}
\def\msbyr{\Msun/\yr}
\def\Rsun{r_{\solar}}
\def\Lsun{L_{\solar}}
\def\Be{{\rm Be}}
\def\Rpole{r_{\ast,p}}
\def\Req{r_{\ast,eq}}
\def\Rmin{r_{\rm min}}
\def\Rmax{r_{\rm max}}
\def\Rstag{r_{\rm stag}}
\def\vinf{V_\infty}
\def\Vrot{V_{rot}}
\def\Vcrit{V_{crit}}
\def\half{{1 \over 2}}
\newcommand{\beq}{\begin{equation}}
\newcommand{\eeq}{\end{equation}}
\newcommand{\bal}{\begin{align}}
\newcommand{\eal}{\end{align}}
\def\phip{{\phi'}}

\maketitle

\begin{abstract}

The extreme luminosities of massive, hot OB stars drive strong stellar winds through line-scattering of the star's UV continuum radiation. 
For OB stars with an orbiting circumstellar disk, we explore here the effect of such line-scattering in driving an {\em ablation }of material from the disk's surface layers,  with initial focus on the marginally {\em optically thin decretion} disks of classical Oe and Be stars.
For this we apply a multi-dimensional radiation-hydrodynamics code that assumes simple optically thin ray tracing for the stellar continuum, but uses a multi-ray Sobolev treatment of the line transfer; this fully accounts for the efficient driving by non-radial rays, due to desaturation of line-absorption by velocity gradients associated with the Keplerian shear in the disk.
Results show a dense, intermediate-speed surface ablation,
consistent with
 the strong, blue-shifted absorption of UV wind lines seen in Be shell stars that are observed from near the disk plane.
A key overall result is that, after an initial adjustment to the introduction of the disk, the asymptotic disk destruction rate is typically just an order-unity factor times the stellar wind mass loss rate. For optically thin Be disks, this leads to a disk destruction time of order months to years, consistent with observationally inferred disk decay times. The much stronger radiative forces of O stars reduce this time to order days,  making it more difficult for decretion processes to sustain a disk in earlier spectral types, and so providing a natural explanation for the relative rarity of Oe stars in the Galaxy.
Moreover, the decrease in line-driving at lower metallicity implies both a reduction in the winds that help spin-down stars from near-critical  rotation, and a reduction in the ablation of any decretion disk; together these provide
a natural explanation for the higher fraction of Classical Be stars, as well as the presence of Oe stars, in the lower metallicity Magellanic Clouds.
We conclude with a discussion of future extensions to study line-driven ablation of denser, optically thick, accretion disks of pre-main-sequence massive stars.

\end{abstract}

 
\section{Introduction}
\label{sec:intro}

In hot, luminous, massive stars,  line scattering of the star's continuum radiation can drive strong stellar wind outflows
\citep{PulVin08}.
The study here focuses on the role of such line-scattering in driving ablation flows from the surface of circumstellar disks around such massive stars.
Such circumstellar disks are observed around both pre-main-sequence (PMS) and main-sequence (MS) massive stars, wherein they respectively act as mediators of mass transfer towards or away from their host star.
The high density of PMS accretion disks means that computation of the line-force must account for optically thick continuum transfer, a complication we leave to a follow-up paper \citep[see chapter 7 of][]{Kee15}.

However, a subpopulation \citep[about 20\%;][]{ZorBri97} of main sequence B-stars, the classical Be stars, show doubled peaked Balmer emission lines thought to arise in a circumstellar {\em decretion} disk, likely formed from orbital ejection of gas from a nearly critically rotating stellar surface \citep[see, e.g.,][for general review of Be stars]{RivCar13}.
Observations indicate such Be decretion disks generally have a moderate density \citep{RivCar13}, with electron scattering continuum radial optical depths in the disk central plane ranging from marginally optically thin ($\tau \lesssim 1$) to modestly optically thick ($\tau \gtrsim 1$).
Since the continuum transfer thus remains optically thin along the bulk of rays impinging  the disk at oblique angles, such Be decretion disks provide a natural testbed for exploring line-driven disk-ablation without the added complication of optically thick continuum transfer. 
Additionally, Be disks are observed to grow and decay on time scales of months to years, making comparisons of models to observations much more tenable than the comparable project for star-forming disks, which persist for timescales of thousands of years or more. 
Finally, as the ubiquitous rapid rotation of classical Be stars is considered to be key to the creation of their disks, these objects provide a natural channel for the extension of this study to include rotational effects.

While observations show $\sim1/5$ of the B stars in our local volume to be Be stars, observations of Oe disks are much less common. In fact, \cite{NegSte04} suggest that the extension of the classical Be phenomena to O-type stars only continues to spectral type O9 with scattered outliers of earlier special type\footnote{The earliest type star they found was HD 155806, an O7.5 star. However, the spectral typing of this, and indeed of all Oe stars, is controversial \citep[see, e.g.][]{VinDav09}.}. Spectropolarimetric observations by \cite{VinDav09} further reinforce the lack of main- or near-main-sequence circumstellar disks around O stars. This break in the appearance of such disks is very evocative and also motivates this work.

\citet{GayOwo99} examined line-driven ablation of a static stellar surface by external radiation, e.g. from a close binary companion.
In the context of cataclysmic variables (CV), several authors  \citep[e.g.][]{FelShl99,Pro99} have modeled disk-winds  driven by a combination of the disk self-luminosiity and line-scattering of radiation from the central white dwarf.
In their study of disk angular momentum transport in Be stars, \citet[][]{KrtOwo11} derived scalings for disk ablation, but using flux scalings that  ignored Keplerian shear.
\citet{GayIgn01} included such shear in their analysis of how line-forces could induce precession of elliptical orbits in the disk, but they did not study disk ablation.

We focus here on ablation of Be disks,  including \citep[as in][]{Pro99} all vector components of the line-acceleration, accounting for velocity gradients arising both from the wind expansion and the disk shear.
We begin in \S \ref{sec:num_phys} by reviewing the relevant background physics including the formalism of line-acceleration and the structure of isothermal, circumstellar gaseous disks. Here we also discuss briefly the effects of rapid rotation on stellar shape through oblateness, and stellar brightness through gravity darkening. \S \ref{sec:results} presents simulations of a standard model of a B2 star, including a discussion of the role of stellar rotation in line-driven ablation. This section also presents our parameter study as a function of spectral type. Finally, in \S \ref{sec:conclusions} we summarize the conclusions of this work as well as address potential lines of future work.

\section{Physics Modules \& Numerical Set-Up}
\label{sec:num_phys}

\subsection{Hydrodynamical Equations}
\label{sec:hydro}
We use the PPM \citep[Piecewise Parabolic Method;][]{ColWoo84} numerical hydrodynamics code\footnote{http://wonka.physics.ncsu.edu/pub/VH-1/} {\tt VH-1} to evolve the time-dependent conservation equations for mass and momentum for the changes in time $t$ of the mass density $\rho$ and vector flow velocity $\bf{v}$,
\beq
 \frac{\partial \rho}{\partial t} +  \nabla  \cdot (\rho \mathbf{v} ) = 0 \, ,
 \label{eq:mcons}
 \eeq
\beq 
 \frac{\partial \mathbf{v}}{\partial t} +  \mathbf{v} \cdot \nabla  \mathbf{v}  =
 - \frac{c_s^2 \nabla \rho }{\rho} 
 - \frac{G M}{r^2} \hat{r}
 + \bf{g}_{\rm lines} .
\label{eq:pcons}
\eeq
For simplicity, we avoid explicit treatment of energy conservation by assuming that the competition between photoionization and radiative cooling maintains a nearly constant gas temperature, taken here to be at the stellar effective temperature, $T=T_{\rm eff}$ \citep[e.g.][]{Dre89}.
This allows us to use the ideal gas law to substitute, within the gas pressure gradient term, for the gas pressure $p = \rho k T/\bar{\mu} = \rho c_s^2$,
with $k$ Boltzmann's constant, $\bar{\mu}$ the mean mass per particle, and latter equality defining the isothermal sound speed $c_s$.
Here $\bar{\mu}$ is calculated assuming full ionization at solar abundances.
The vector equations are evolved in spherical coordinates $\{r,\theta,\phi\}$, assuming axial symmetry in the ignorable coordinate $\phi$.

The right side of equation (\ref{eq:pcons}) also includes the vector radiative acceleration from line scattering $\bf{g}_{\rm lines}$ (discussed further in the next subsection, \S \ref{sec:grad}), and the inward radial stellar gravity\footnote{We ignore self-gravity of the circumstellar material, as it is here many orders of magnitude weaker than that of the central star.} at radius $r$, with $G$ Newton's gravitation constant;
here the effective stellar mass
$M \equiv M_\ast(1-\Gamma_e)$ accounts for the reduction in effective gravity by continuum electron scattering opacity $\kappa_e (= 0.34$\, cm$^2$\,g$^{-1}$), through the ratio of radiative to gravitational acceleration,
\beq
\Gamma_e \equiv\frac{\kappa_e L_\ast}{4 \pi G M_\ast c} \, .
\eeq

\subsection{Line Acceleration}
\label{sec:grad}

For the line-acceleration, we follow the methodology of \cite{CasAbb75} (hereafter CAK) in assuming a power-law distribution in line strength, cast here in terms of the ``quality'' of the line resonance \citep{Gay95},
\beq
q\equiv \frac{\kappa_L}{\kappa_e}\frac{v_{\rm th}}{c}\, ,
\eeq
with $v_{\rm th}$ the ion thermal speed, $c$ the speed of light, and $\kappa_L$ the line-integrated opacity (measured in units of cm$^2$ g$^{-1}$).
In this notation, the differential number distribution in line-strength is given by
\beq\label{eq:line_ens}
\frac{dN}{dq}=\frac{\bar{Q}}{\Gamma(\alpha) Q_\mathrm{o}^2}\left(\frac{q}{Q_\mathrm{o}}\right)^{\alpha-2}e^{-q/Q_\mathrm{o}}\, ,
\eeq
where $\alpha$ is the power-law index introduced by CAK. 
Here, the pure CAK power-law distribution is truncated at a maximally strong line of enhancement $Q_\mathrm{o}$ over continuum opacity, while the normalization $\bar{Q}$ represents the total opacity enhancement in the limit when all the lines are optically thin.
For stars with effective temperatures $T_{\rm eff} \gtrsim 30$ kK, $Q_\mathrm{o} \approx \bar{Q}$ \citep[e.g.][]{Gay95}. 
However, as shown by the NLTE calculations of \cite{PulSpr00}, for cooler B stars, $\bar{Q}$ and $Q_\mathrm{o}$ can differ by factors of a few.
Therefore we here use values of $Q_0$ and $\bar{Q}$ derived by \cite{PulSpr00}. 
These line quality parameters are related to the standard CAK line-force normalization by
\beq
k_{\rm CAK}=\frac{1}{1-\alpha}\left(\frac{v_{\rm th}}{c}\right)^\alpha \bar{Q}Q_\mathrm{o}^{-\alpha}\, .
\eeq

For a radially accelerating, expanding outflow, all photons only interact with a spectral line at a single resonance location, where the frequency of the photon has been Doppler shifted into the line resonance \citep{Sob60}. 
For a photon in direction $\hat{\mathbf{n}}$, the Sobolev optical depth across this resonance is, for a line of opacity enhancement q, given by 

\beq
\tau_\mathrm{q}(\hat{\mathbf{n}})\equiv \frac{\rho \kappa_e q c}{dv_n/dn}\, ,
\eeq
with the line-of-sight velocity gradient in the $\hat{\mathbf{n}}$ direction given by  $dv_n/dn\equiv \hat{\mathbf{n}}\cdot\nabla(\hat{\mathbf{n}}\cdot\mathbf{v})$.
The associated enhancement of the line-acceleration over pure electron scattering is then

\beq
\frac{g_{\rm line}}{g_e}=q\frac{1-e^{-\tau_q}}{\tau_q}\,.
\eeq
For an ensemble of isolated (i.e. non-overlapping) lines, integration over the distribution (\ref{eq:line_ens}) and the full solid angle of the star with specific intensity $I_\ast (\hat{\mathbf{n}} )$ gives for the total line-acceleration,
\beq\label{eq:g_line}
\mathbf{g}_{\rm lines}=\frac{\kappa_e \bar{Q}}{(1-\alpha) c} \oint \left[\frac{(1+\tau_\mathrm{o}(\hat{\mathbf{n}}))^{1-\alpha}-1}{\tau_\mathrm{o}(\hat{\mathbf{n}})}\right]\,  I_\ast (\hat{\mathbf{n}}) \hat{\mathbf{n}}\, d\Omega\,,
\eeq
where $\tau_\mathrm{o}$ is $\tau_\mathrm{q}$ for $q=Q_\mathrm{o}$. 
This is the general, 3-D vector form used for the line-acceleration calculations in this paper.

In the limit when the strongest line is very optically thick (i.e. $\tau_\mathrm{o}\gg1$), the factor in square brackets reduces to $\tau_\mathrm{o}^{-\alpha}$, giving

\beq\label{eq:g_line_thick}
\mathbf{g}_{\rm lines}\approx\frac{\kappa_e \bar{Q}}{(1-\alpha)c(Q_\mathrm{o} \kappa_e c \rho)^\alpha}\oint\left(\hat{\mathbf{n}}\cdot\nabla(\hat{\mathbf{n}}\cdot\mathbf{v})\right)^{\alpha}\, I_\ast (\hat{\mathbf{n}}) \hat{\mathbf{n}}\, d\Omega\, .
\eeq
For the classical CAK model with a point star, and so radially streaming photons, the line acceleration is purely radial with a magnitude, $g_{CAK}\propto (dv_r/dr)^\alpha$, that depends on the radial velocity gradient, $dv_r/dr$. In the standard CAK formalism, this line acceleration gives rise to the radial velocity gradients that sustain the self-consistent wind solution.

\begin{figure}
\includegraphics[width=0.5\textwidth]{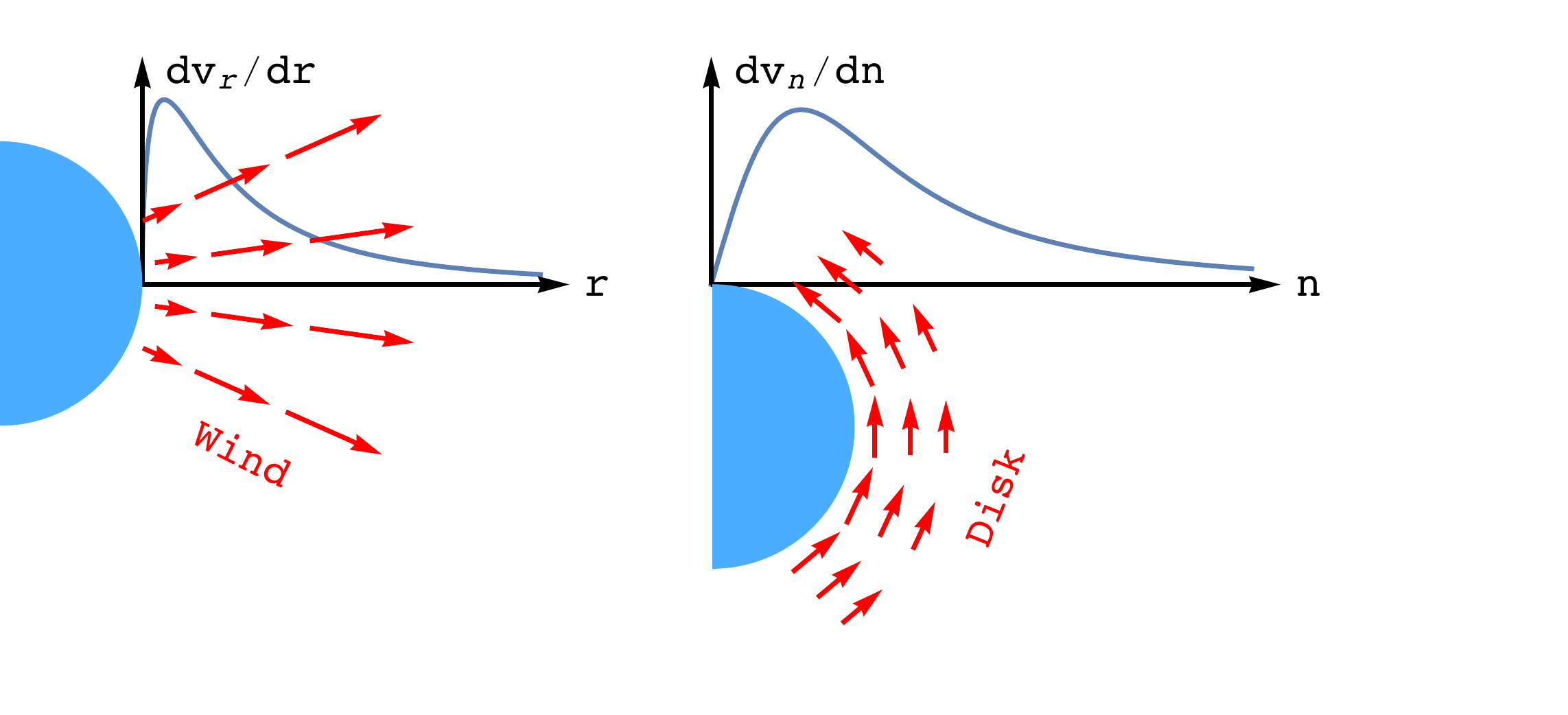}
\caption{
Schematic depiction of the velocity gradient in a radially expanding wind, $dv_r/dr$ (left), and along a non-radial ray direction $n$ through a Keplerian disk with azimuthal shear, $dv_n/dn$ (right).
}
\label{fig:vel_grad}
\end{figure}

For an orbiting circumstellar disk, the radial velocity is effectively zero everywhere, implying that the radially streaming photons should impart no line force. 
However, as illustrated in figure \ref{fig:vel_grad}, the inherent Keplerian shear in the azimuthal orbital velocity of the disk, which declines in radius as $v_\phi\propto 1/\sqrt{r}$, means that non-radial photons see a line-of-sight velocity gradient that desaturates the line-absorption, and so gives rise to a net line force. 
In the relatively low-density surface layers, the associated acceleration can overcome gravity and lead to a substantial ablation of material from the disk.

\begin{figure}
\centering
\includegraphics[width=0.45\textwidth]{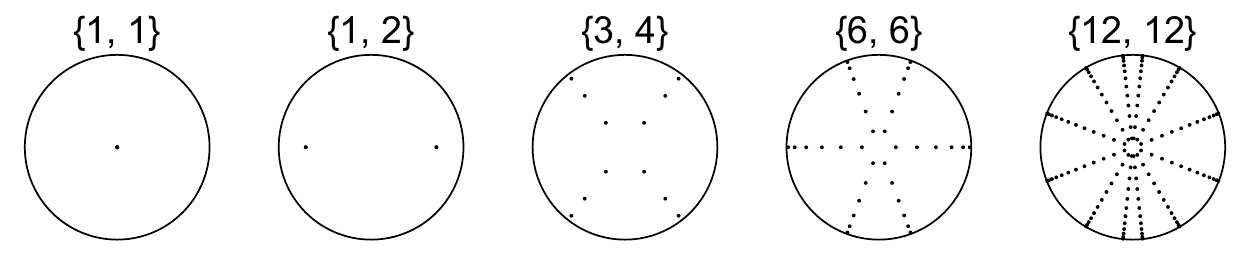}
\caption{
Characteristic ray quadratures for resolving the stellar disk. Each label lists first the number of impact parameters $p$,  and then the number of azimuthal angles $\phi'$. This work uses the 6$\times$6 quadrature.
}
\label{fig:ray_quad}
\end{figure}

To accurately quantify this in a dynamical simulation, it is necessary to account fully for the illumination of the disk by non-radial radiation from the finite angular size of the star. 
As illustrated in figure \ref{fig:ray_quad}, we have experimented with a range of numerical quadratures for rays that intersect the star with various impact parameters $p$ from the center,  and with various azimuthal angles $\phi'$, as seen from a given point in the wind. The points on the star are distributed using a Gauss-Legendre quadrature in $p^2$ and $\phi'$. The rays joining each of the points on the stellar core with the wind reference point then give the line of sight directions for calculating the velocity gradients needed in equation (\ref{eq:g_line}). The details of the selection of ray quadrature are discussed further in section \ref{sec:num}.

In following this standard CAK-Sobolev approach originally developed for a spherically expanding wind, we are assuming that radiation from the stellar core has a single line-resonance at any local point where we are computing the line-acceleration.
However, as discussed in Appendix \ref{sec:appa}, in a configuration with a fast wind at high latitudes and slow or no radial outflow in the equatorial disk, core rays impinging the disk from above and below can sometimes have a {\em non-monotonic} line-of-sight speed variation, implying then the possibility of {\em multiple} line resonances.

Any such resonance between the disk viewpoint and the stellar surface can then absorb or scatter the radiation from the stellar core, reducing the local radiative illumination and thus the line-driving of the disk.
Moreover, accounting for such ``inner-resonance shadowing'' requires in general a global solution of the non-local coupling among all such resonant scattering points, something that would be computationally very costly to incorporate into a hydrodynamical simulation of line-driven disk ablation.

Fortunately, the analysis in Appendix  \ref{sec:appa} indicates that, for the continuum optically thin Be disks considered here, the net reduction in line-driving is likely to be less than a factor two.
For computational tractability in this initial study, we thus choose simply to ignore such multiple line resonance effects.

\subsection{Stellar Rotation}
\label{sec:star_rot}

The quadratures above assume the star is spherical, but for rapidly rotating Be stars there can be significant oblateness and associated equatorial gravity darkening.
To account for oblateness, we use the formulation of \cite{ColHar66}, with the standard gravity darkening given by \cite{von24}, both of which are succinctly reviewed by \cite{Cra96}.
The specific rotating model considered has an equatorial surface rotation speed that is 80\% of the near surface equatorial orbital speed.
This leads to an oblateness characterized by an equatorial to polar radius ratio $\Req/\Rpole=1.32$, with an associated ratio of equatorial to polar flux of $F_{eq}/F_{pole}=0.2$. 
The rotating models have their surface-integrated flux normalized such that total stellar luminosity is preserved.
To enforce this, the quadratures discussed in the previous section are remapped onto an oblate spheroid by numerically determining the angular extent of the star along each of the $\phi'$ directions used, and then distributing the points in impact parameter over this range.
Each ray is then assigned a flux weighting proportional to the effective gravity at its stellar footpoint.
As the simulation is isothermal, the associated $\theta$ dependence of stellar effective temperature is ignored.
For simplicity, the models presented here also ignore limb darkening, but tests we have done show that adding limb darkening leads to less than a 10\% reduction in disk mass ablation rate.

\subsection{Disk Initial Condition}
\label{sec:disk_phys}

To set the initial conditions for our simulations, we first relax a time-dependent wind simulation to a steady state for each case. 
We then superimpose a simple analytic model of a circumstellar disk, now specified in cylindrical coordinates $\{R,z,\phi\}$, with $R\equiv r \sin \theta$, $z\equiv r \cos \theta$, and again assuming axial symmetry in the ignorable coordinate $\phi$.
As reviewed by \cite{Bjo97}, a geometrically thin, isothermal disk whose vertical extent is controlled by the competition of gas pressure and stellar gravity has a Gaussian vertical structure such that
\beq
\rho(R,z)=\rho_{eq}(R)e^{-z^2/2H(R)^2}
\eeq
where the local scale-height is given by
\beq
H(R)=\frac{c_s}{v_{orb}(R_\ast)} \frac{R^{3/2}}{R_\ast^{1/2}}\, ,
\eeq
where $R_\ast \equiv \Req$, and the density in the equatorial plane at cylindrical radius $R$ is assumed to follow a power-law,
\beq
\rho_{eq}(R)=\rho_\mathrm{o}\left(\frac{R}{R_\ast}\right)^{-n}.
\label{eq:disk_struc}
\eeq
This choice is supported for Be disks by the power-law+Gaussian-disk radiative transfer models of \cite{TouGie11}. A systematic fitting by \cite{SilJon10} of H$\alpha$ profiles from the decretion disks of classical Be stars shows that $n$ can vary between\footnote{For pre-main sequence accretion disks the power-law index is found by \cite{FisHen96} to be between 0.5 and 2.} 1.5 and 4, with statistically significant peak aroung $n \approx  3.5$.  
Since this is also the index found analytically for viscous transport in a decretion disk \citep{RivCar13}, and is the favored choice of fixed $n$ for other classical Be models \citep[e.g.][]{CarOka09}, we adopt here $n=3.5$ for all models.

By taking the steady state form of equations (\ref{eq:mcons}) and (\ref{eq:pcons}), and the density structure of the disk from equation (\ref{eq:disk_struc}), the approximate azimuthal velocity neccessary to centrifugally support the disk is
\beq
v_\phi(R,z)=\sqrt{\frac{G\Mstar}{R}}\frac{|z|}{\sqrt{R^2+z^2}}\sqrt{1 + n\left(\frac{H}{R}\right)^2}\, .
\label{eq:disk_v_phi}
\eeq
Here the second term in the radical represents a small radial pressure correction to the standard form of Keplerian orbital velocity, included to create initial conditions as close as possible to a stable equilibrium state.

For opacity arising purely from electron scattering, the radial optical depth through the disk midplane is

\beq
\tau_{disk}=\int_{\Rstar}^\infty \kappa_e \rho_{eq} dR =  \frac{\kappa_e \rho_\mathrm{o} R_{eq}}{n-1}\, .
\eeq
To ensure the disks remain marginally optically thin, we set the base density of all models to $\rho_\mathrm{o}=1/(\kappa_e R_{eq})$, which gives $\tau_{disk}=1/(n-1)=0.4$. For the fully ionized, solar metallicity value $\kappa_e=0.34$ cm$^2$ g$^{-1}$ and $R_{eq}=5R_\odot$ (as used below for the standard B2 model), $\rho_\mathrm{o} = 8.5 \times 10^{-12}$ g cm$^{-3}$, within the range of typical values inferred for classical Be stars \citep[see, e.g.][]{RivCar13}.

\begin{figure}
\centering
\includegraphics[width=0.48\textwidth]{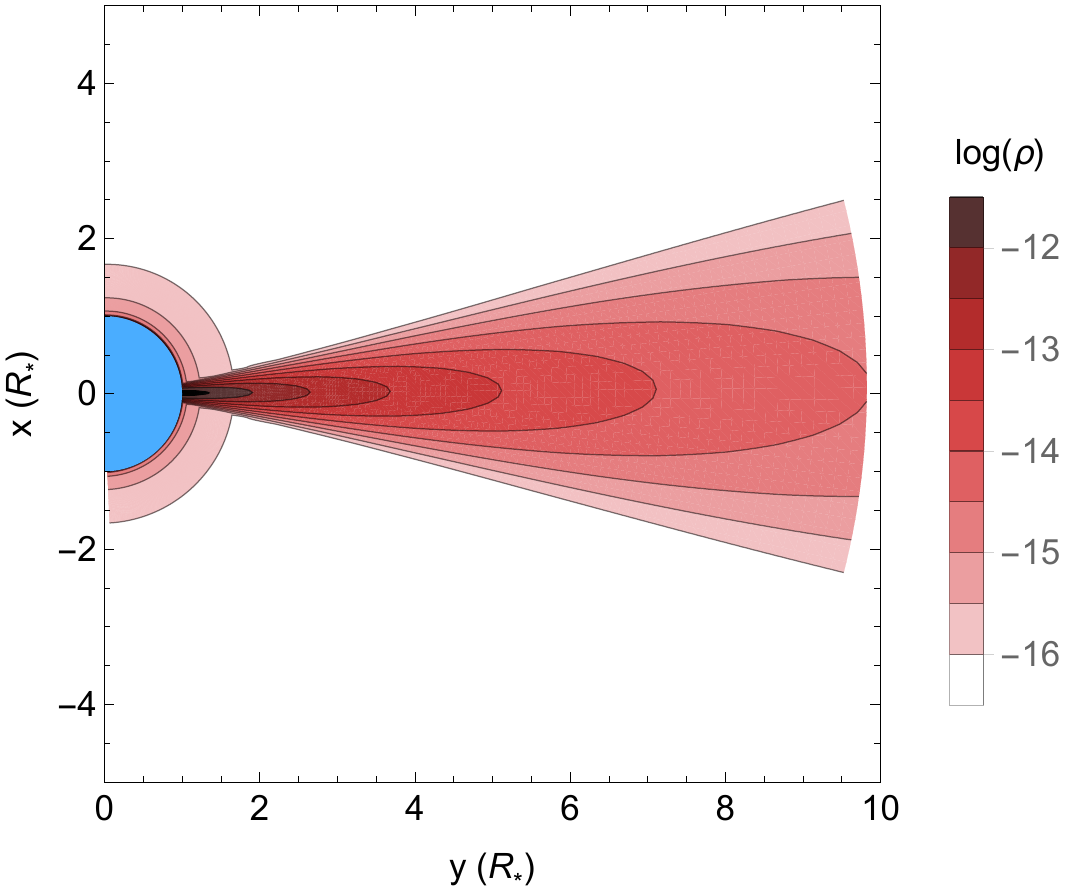}
\caption{
Contours for the spatial distribution of log density (in g\,cm$^{-3}$) for the
initial condition of our B2e standard model.
}
\label{fig:b2_init}
\end{figure}

Using this base density for the disk, figure \ref{fig:b2_init} plots contours for the spatial distribution of $\log(\rho)$ for the initial condition of our standard model of a B2e star.

\subsection{Numerical Specifications}
\label{sec:num}

The simulations here must resolve both the spatial variations  of density and velocity in $r$ and $\theta$ (or $R$ and $z$), as well the stellar intensity as a function of impact parameter $p$ and observer centered angular position on the star $\phi'$. 

As summarized in Table \ref{tab:grid}, the spatial grid ranges from pole to pole in latitude, and over 1-10 $\Req$ in radius, with  the highest latitudinal resolution at the equator (to resolve the hydrostatic disk), and the highest radial resolution near the surface (to resolve the steep transition to transonic wind over about an atmospheric scale height).
Away from the disk or the stellar surface, the respective grid spacings are increment across each successive zone by the fixed  multiplicative ``stretch factors" quoted in Table \ref{tab:grid}.

\begin{table}
\centering
\caption{Grid specifications \label{tab:grid}}
\begin{tabular}{| l | c  c |}
\hline
& $r$ & $\theta$  \\
\hline
min. & $R_\ast$ & 0\\
max. & $10\,R_\ast$ & $\pi$ \\
number of zones & 300 & 120 \\
stretch & 1.02 & 1.015 \\
min. grid spacing & 4.7$\times 10^{-4}\,R_\ast$ & 1.7$\times 10^{-2}\, R_\ast$ \\
\hline
\end{tabular}
\end{table}

For the ray quadrature,  ensuring sufficient resolution is less apparent, largely due to the fluctuating small scale structures generated by ablation. Since each viewing direction shows a different velocity gradient, these small scale structures can be of different magnitudes and, in some cases, even different positions with changing ray quadratures. The sharp interface between the disk and wind also contributes to significant differences between velocity gradients for neighboring rays. On a grid level, it is thus difficult to identify a quadrature with clear formal convergence, so a more global criterion is needed. Referring to the testing in \cite{Kee15}, we find that a quadrature with 6 points in both impact parameter and $\phi'$ is sufficiently dense to ensure that the quantities of interest, most importantly the rate of disk ablation, are not a artifact of the chosen quadrature.

For boundary conditions at the poles we enforce standard axisymmetry.
At the stellar base, we assume a fixed density\footnote{This base density is $\rho_\ast=5\rho_{\rm sonic}$, with $\rho_{sonic}\equiv\dot{M}_{CAK}/(4\pi r_{\rm sonic}^2 c_s)$ the density at the sonic radius $r_{\rm sonic}$, where $v_r(r_{\rm sonic})=c_s$. Here $\dot{M}_{CAK}$ refers to the analytic CAK mass loss rate including the finite disk correction factor. For this B2 model, $\rho_\ast=5\times10^{-14}$g/cm$^3$.}
independent of latitude, and set the radial velocity by linear extrapolation, with the constraint that $|v_r| \leq c_s$.
We also impose zero latitudinal velocity, and azimuthal velocity set by the stellar rotation at this boundary.
At the outer boundary, we use a simple extrapolation assuming a zero radial gradient in all quantities.
This allows for simple outflow in the wind regions, while keeping any outer oscillations in the disk to a small amplitude.

\section{Results}\label{sec:results}

\subsection{Model of a non-rotating B2 star}\label{sec:results_B2}

For our standard model, we choose a B2 star both because of its moderately strong luminosity and also because this is near the spectral type with the largest fraction of Be stars.
Table \ref{tab:b2_params} provides the full set of stellar and disk parameters as derived from the evolutionary tracks of \cite{GeoEks13} and the effective temperature calibrations of \cite{TruDuf07}. Table \ref{tab:b2_wind} gives the wind parameters as derived by \cite{PulSpr00}.

\begin{table}
\caption{Stellar and Disk Parameters of the B2 Standard Model} \label{tab:b2_params}
\centering
\begin{tabular}{| c | c | c | c | c |}
\hline
$T_{\rm eff}$ (kK) & $L_\ast$ ($L_\odot$) & $M_\ast$ ($M_\odot$) & $R_\ast$ ($R_\odot$) & $M_{disk}$ ($M_\odot$)\\
\hline
22 & 5.0$\times 10^3$ & 9 & 5.0 & 1.9$\times 10^{-10}$  \\
\hline
\end{tabular}
\end{table}

\begin{table}
\caption{Wind Parameters of the B2 Standard Model} \label{tab:b2_wind}
\centering
\begin{tabular}{| c | c | c | c |}
\hline
$\bar{Q}$ & $Q_\mathrm{o}$ & $\alpha$ & $\dot{M}_{wind}$ ($M_\odot/yr$) \\
\hline
1800 & 4900 & 0.59 & 5.8$\times 10^{-10}$ \\
\hline
\end{tabular}
\end{table}

\begin{figure*}
\centering
\begin{subfigure}[b]{0.5\textwidth}
\includegraphics[width=\textwidth]{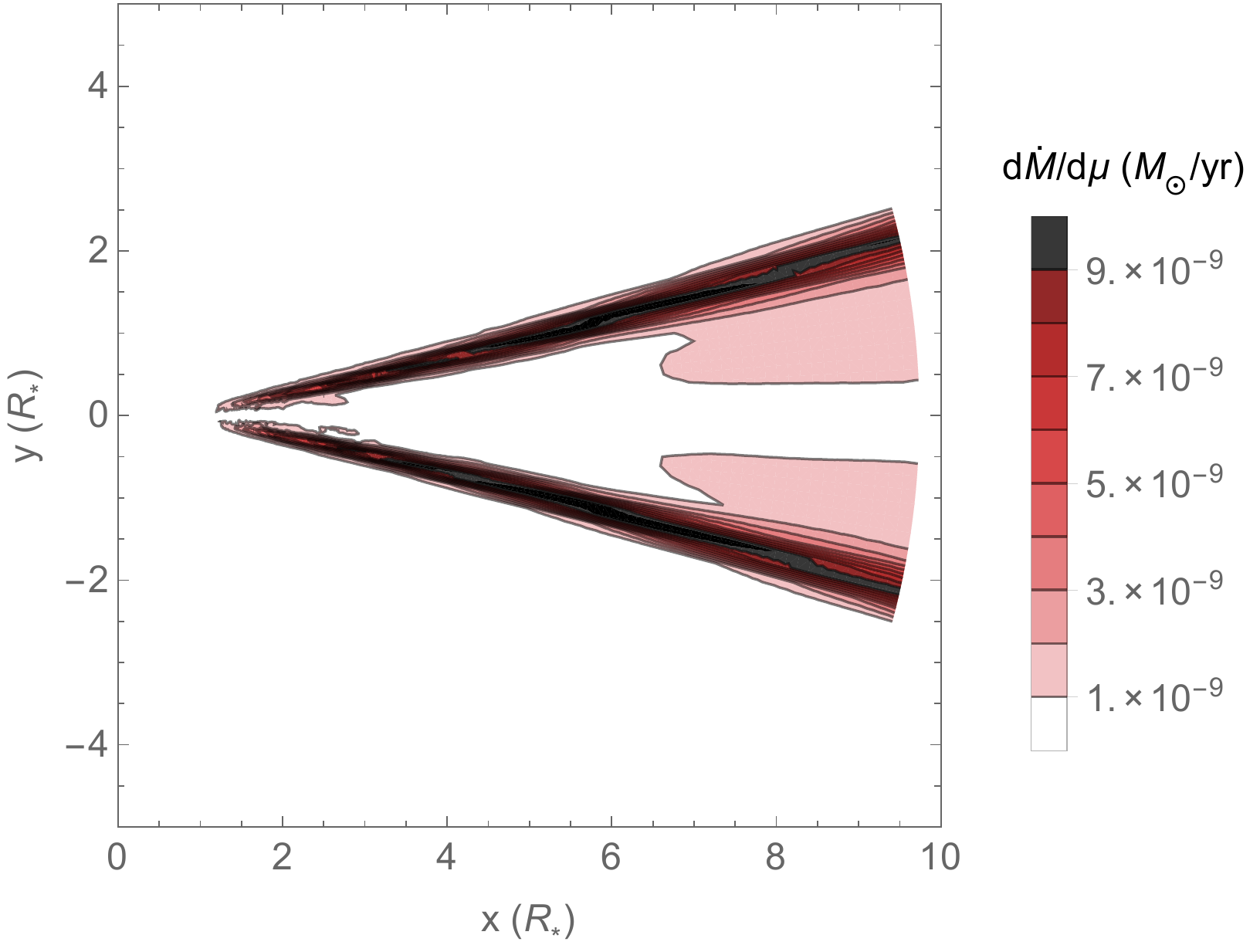}
\end{subfigure}
\begin{subfigure}[b]{0.48\textwidth}
\includegraphics[width=\textwidth]{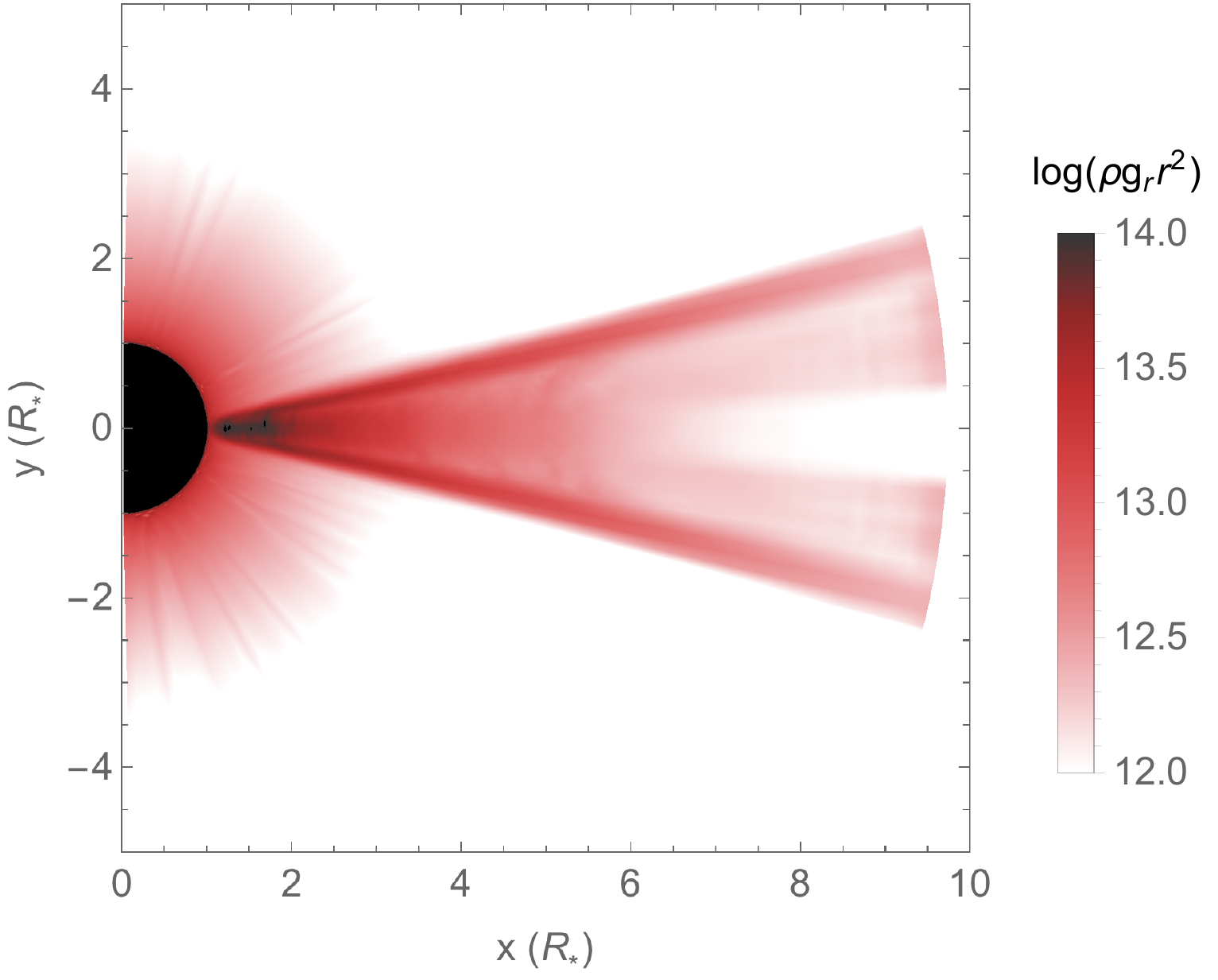}
\end{subfigure}
\caption{
{\em Left:} Latitudinal distribution of mass loss $d{\dot M}/d\mu$ (equation \ref{eqn:dmdotdmu}) for the standard B2 model, plotted in units of $M_\odot$ yr$^{-1}$. 
{\em Right:} Force-per-unit-length $\rho g_r r^2$ in cgs units. Both quantities have been time averaged from $1\times10^6$ to $3\times10^6$ s to omit the initial interval of disk readjustment to the introduction of radiative forces.
}
\label{fig:B2_dmdotdmu}
\end{figure*}

For this non-spherical but azimuthally symmetric simulation, the latitudinal distribution of mass loss can be characterized by

\beq
\frac{d\dot{M}}{d\mu} \equiv 2 \pi \rho v_{r} r^2\,,
\label{eqn:dmdotdmu}
\eeq
where $\mu\equiv\cos\theta$.
The left panel of figure \ref{fig:B2_dmdotdmu} plots the time average of this quantity from $t=10^6$ to $3\times10^6$ s, well after the adjustment of the simulation from its initial conditions.
The results vividly illustrate that ablation  mostly occurs in thin layers along the upper and lower edges of the disk\footnote{The small pink excursions into the central region of the disk are the remnants of small scale oscillatory motions set off by the relaxation of the initial conditions. Elsewhere in the disk, these have averaged out.}. 

To illustrate the dynamical driving associated with this ablation, the right panel of figure \ref{fig:B2_dmdotdmu} plots the log of the radial force-per-unit-radius, $\rho g_{\rm r} r^2$. Note that the ablated material is accelerated throughout the full radius, but since the opening angle of the layer remains nearly constant in radius, we interpret this as
continuing acceleration of the material that has been dislodged from the disk near the star, rather than an addition of new disk material to the ablation flow. 

\begin{figure}
\centering

\begin{subfigure}[b]{0.48\textwidth}
\includegraphics[width=\textwidth]{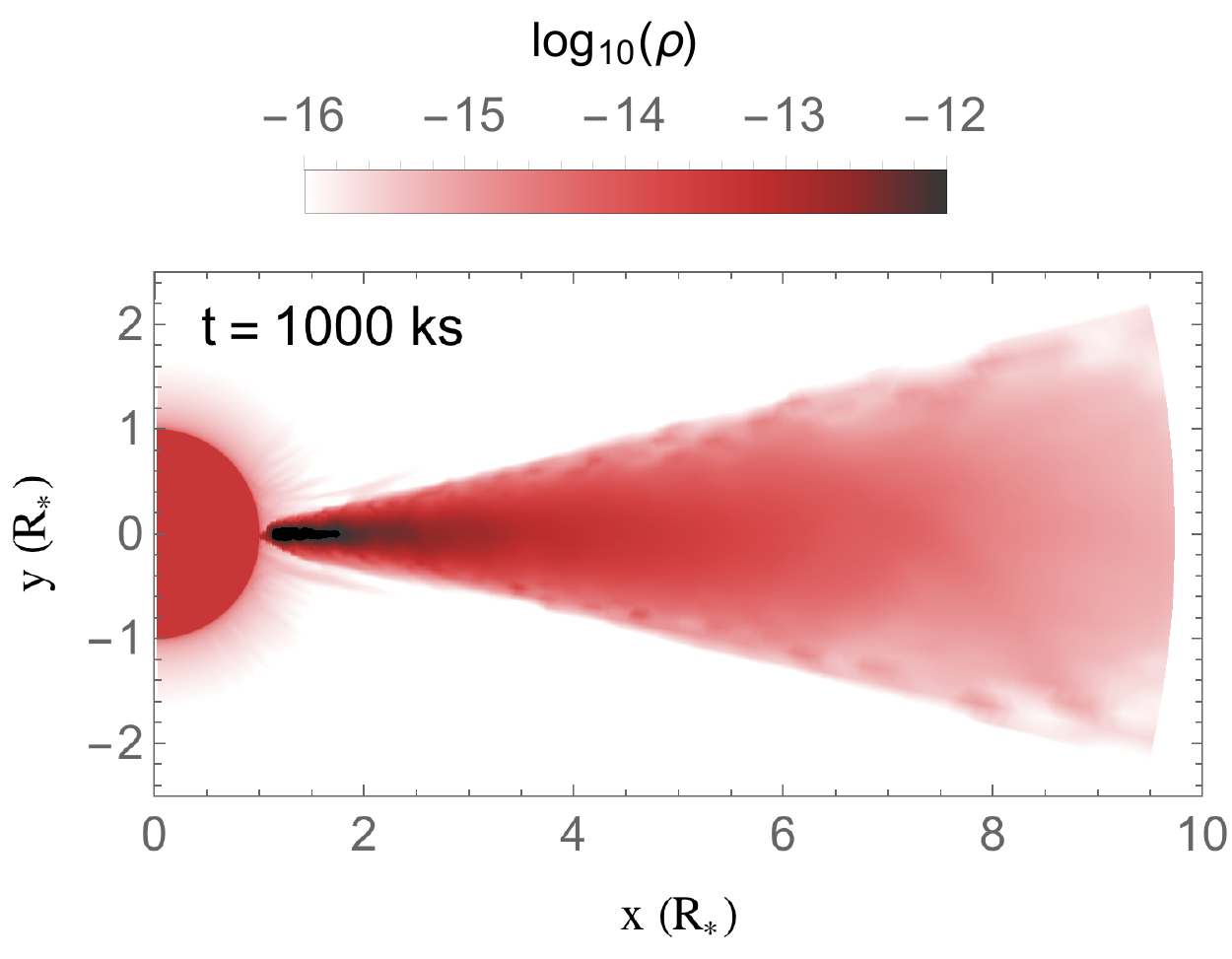}
\end{subfigure}

\begin{subfigure}[b]{0.48\textwidth}
\includegraphics[width=\textwidth]{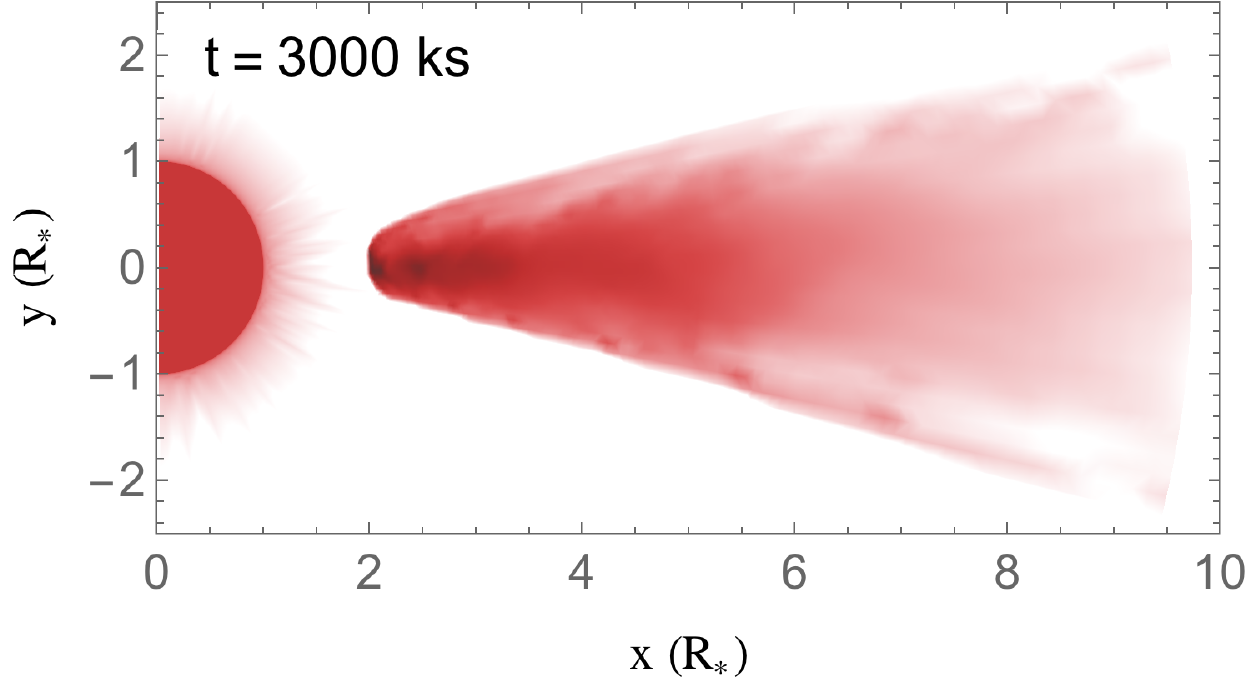}
\end{subfigure}

\caption{
Snapshots of log density for the B2 model at times $t=1000$ and 3000 ks from the initial condition, showing the removal of the disk from the stellar surface outwards.
}
\label{fig:B2_rho_snaps}
\end{figure}

\begin{figure}
\centering
\includegraphics[width=0.5\textwidth]{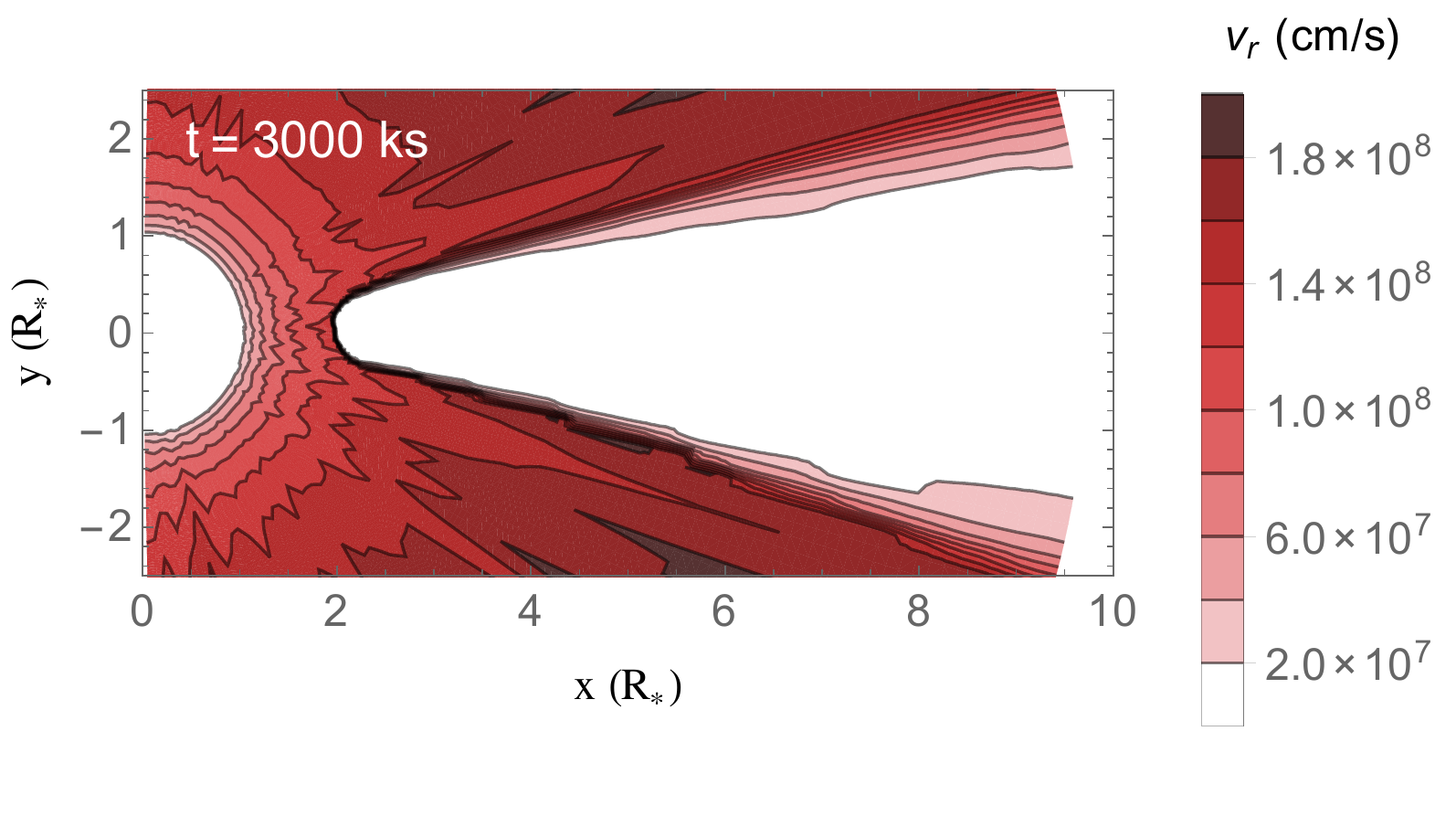}
\caption{
Snapshot of radial velocity for the B2 model at time $t=3000$ ks from the initial condition, emphasizing the thinness of the slow-moving ablation layers, with $v_r\sim 100$ km s$^{-1}$.
}
\label{fig:B2_vel_snap}
\end{figure}

This interpretation is supported by comparison of density snapshots from each model as a function of time. As  illustrated in figure \ref{fig:B2_rho_snaps}, the disk is removed from its inner edge outwards. Additionally,  as shown in figure \ref{fig:B2_vel_snap}, the radial velocity is negligible in the bulk of the disk, but reaches a few hundred km s$^{-1}$ within thin wind-disk boundary layers corresponding to those shown in figure \ref{fig:B2_dmdotdmu}. Observationally, this implies emission lines of Classical Be stars should decay most rapidly in their outer wings, where emission comes from the highest orbital velocities near the stellar surface. It also suggests that the Doppler shift with respect to line center of Classical Be star emission lines should remain approximately constant throughout the decay, as this is set by the outer disk edge, which remains relatively unperturbed throughout much of the ablation process.

\begin{figure}
\centering
\includegraphics[width=0.48\textwidth]{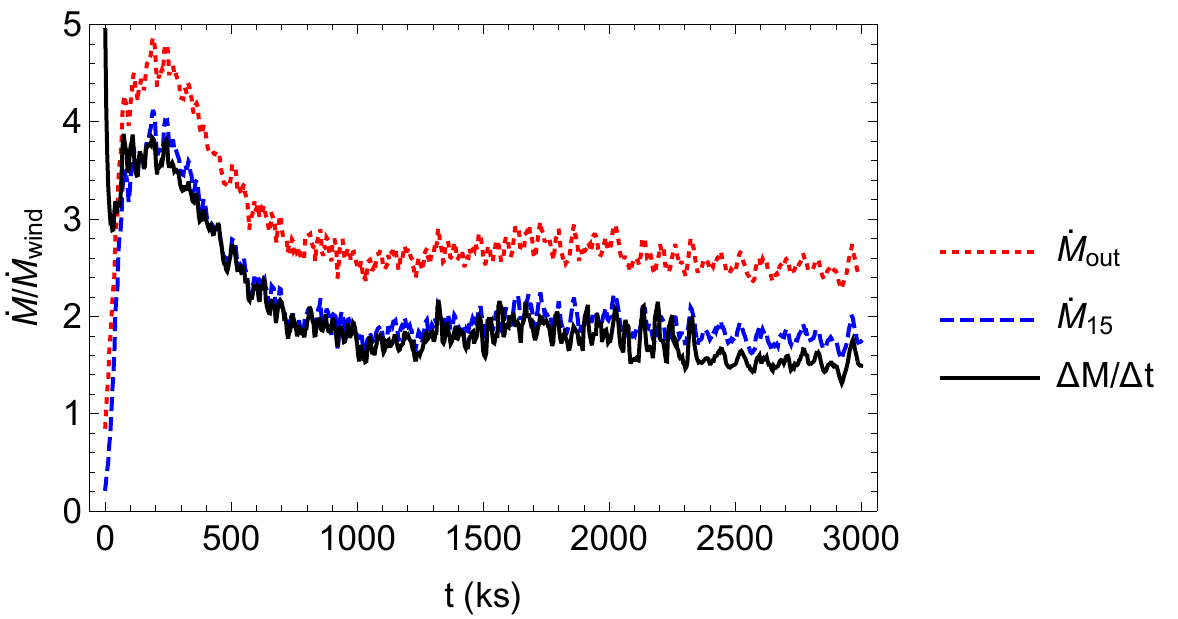}
\caption{
Mass loss rate (win units of spherically symmetric wind mass loss rate) for three mass loss metrics of the standard model of a B2 star.
}
\label{fig:B2_mdot}
\end{figure}

While the disk ablation is strong in thin surface layers of the disk, when integrated over solid angle the associated mass loss is only a factor of a few higher than that of the steady-state wind.
Figure \ref{fig:B2_mdot} quantifies this by plotting three different characterizations of the simulation mass loss, normalized by the steady-state wind rate. 
The uppermost curve, labeled $\dot{M}_{out}$, is the total mass loss rate through the outer boundary, while $\dot{M}_{\rm 15}$ represents the outer boundary mass loss in a band of $\pm 15^\circ$ above and below the equator. 
The latter is quite comparable to the total rate of change of mass in the simulation\footnote{The initial spike in the rate of change of mass in the simulation is due to mass falling back onto the star during the adjustment phase.}, $\Delta M/\Delta t$, indicating that both these are good representations of the total disk ablation rate, with little contribution from the steady wind. After an initial adjustment, the disk ablation rate settles down to about a factor of two times the wind mass loss rate. The small difference toward the final time between $\dot{M}_{\rm 15}$ and $\Delta M/\Delta t$ can be accounted for by the shrinking of solid angle of the disk and the contamination of $\dot{M}_{\rm 15}$ by wind mass loss. In the following we use $\Delta M/\Delta t$ as the best characterization of disk ablation rate.

\subsection{Importance of non-radial velocity gradients}

\begin{figure}
\centering
\includegraphics[width=0.5\textwidth]{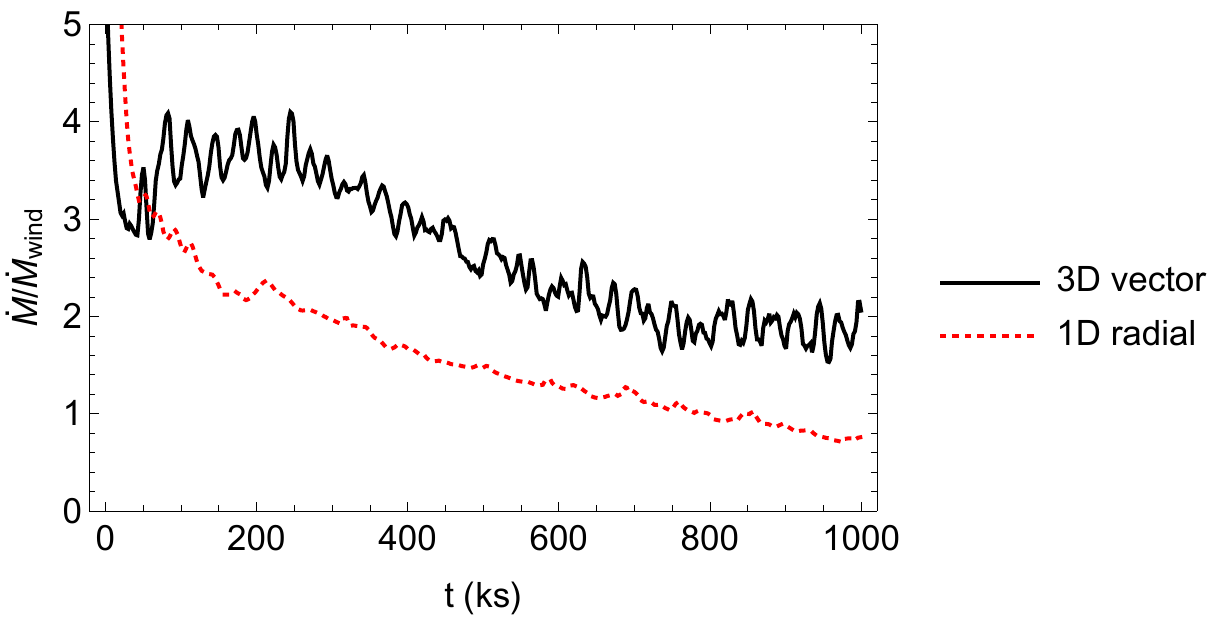}
\caption{
Ablation rate, measured in units of the spherically symmetric mass loss rate, for the two force implementations discussed in the text.
}
\label{fig:1d_v_3d}
\end{figure}

While the full results presented in the rest of this paper use the complete 3D vector acceleration described in \S \ref{sec:grad}, it is instructive for testing to examine reduced models that assume 
 $\mathbf{v}(r)=v_r\mathbf{\hat{r}}$  and $v_\theta=v_\phi=\partial v_r/\partial \theta = \partial v_r/\partial \phi=0$ in computing a purely radial\footnote{This is achieved by replacing $\mathbf{\hat{n}}\cdot\nabla(\mathbf{\hat{n}}\cdot\mathbf{v})$ with $\mu^2 dv_r/dr+ (1-\mu^2) v_r/r$ in equation \ref{eq:g_line_thick}.} 
line-acceleration $\mathbf{g}=g_r\mathbf{\hat{r}}$. 
Since such conditions do apply in the high-latitude wind,
we expect the wind to be unchanged among the three implementations, and this comparison can test the importance of the non-radial velocity gradient terms in causing ablation. It can also help disentangle what portion of the mass lost from the disk is due to ablation -- by which we mean the removal of material only by direct acceleration of disk material by radiation -- and what fraction is from ``entrainment'', wherein the wind drags away low density disk material viscously coupled through a Kelvin-Helmholtz instability. Figure \ref{fig:1d_v_3d} shows that, asymptotically, the 3D vector form yields a disk ablation rate more than twice the 1D radial model. Thus, we can infer that, while the strong velocity shear between wind and disk does indeed entrain material, this contributes at most half of the total disk ablation rate.

\subsection{Sample comparison 
 with {\tt IUE} spectrum of Be shell star}
 \label{sec:shelluvcomp}

\begin{figure}
\centering
\includegraphics[width=0.48\textwidth]{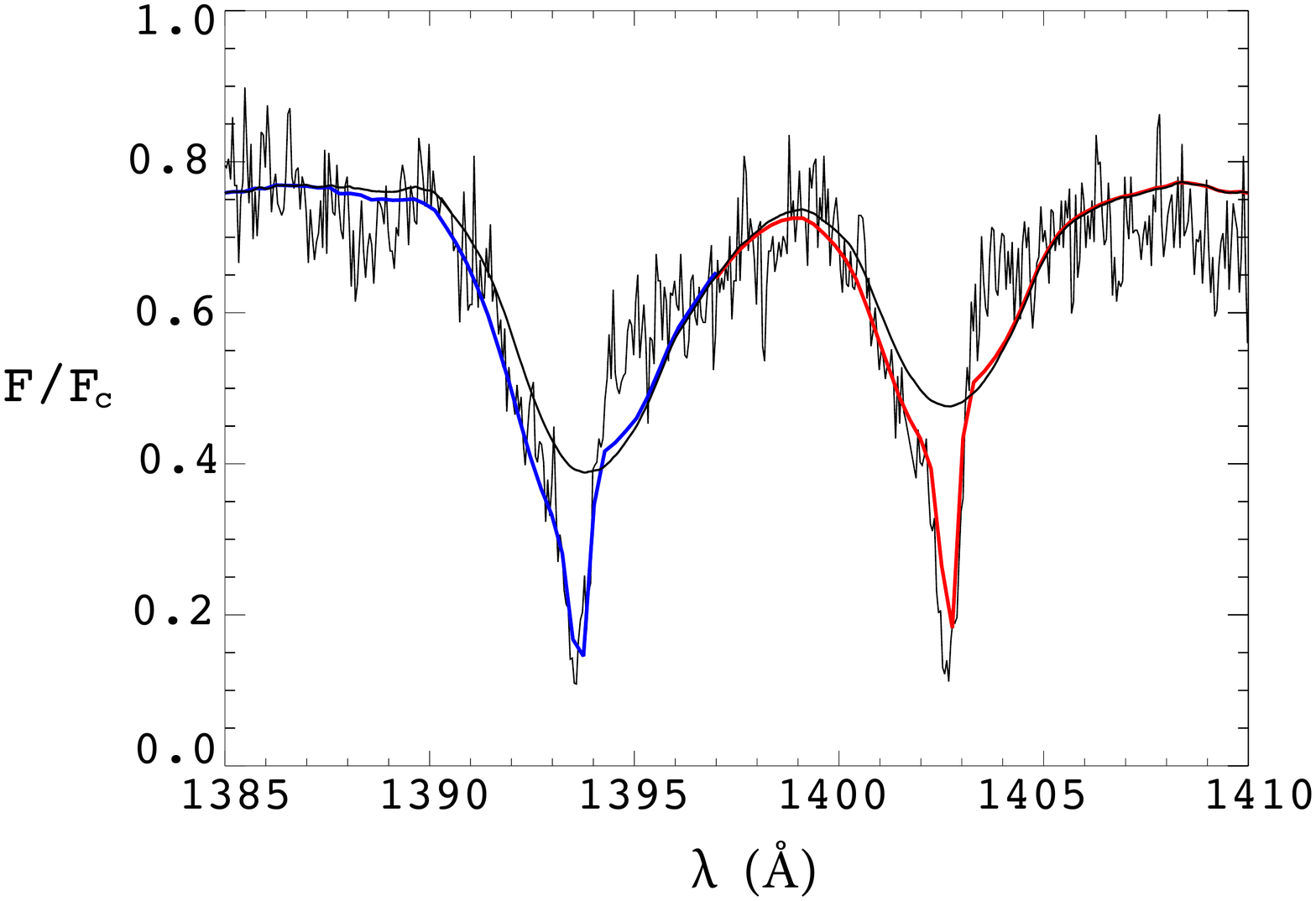}
\caption{
{\tt IUE} flux spectrum for the  Si\,IV line doublet  of the Be shell star $\phi$\,Per,  overplotted vs.\ wavelength (in \AA)  with synthetic line profiles.
The black curve
showing broad, shallow absorption 
is for photospheric profiles from {\tt TLUSTY} \citep{LanHub07} for the star's spectral type 
\citep[B2Ve;][]{Les68}, rotationally broadened according to the star's Vsini=450\,km\,s$^{-1}$ \citep{AbtLev02}.
The deeper, narrow absorption troughs show the blue/red components of the Si\,IV doublet, computed from applying a UV line-synthesis code for our B2 disk ablation model, viewed from 6 degrees above the equator.
The {\tt IUE} continuum is reduced by a factor 0.8 to account for line blanketing in the synthetic pseudo-continuum.
}
\label{fig:shell}
\end{figure}

This strong concentration of mass ablation along the surface layers of the disk seems qualitatively what is needed to explain the blue-shifted absorption troughs seen in {\tt IUE} spectra of UV wind lines (e.g.\ Si\,IV) from Be shell stars that are observed from near the equatorial plane \citep{GraBjo87}.
Preliminary calculations we have performed with a multi-D UV line synthesis code described in \citet{NazSun15} indeed show that these models can match quite well  UV line profiles from {\tt IUE} spectra of specific Be shell stars; figure \ref{fig:shell} shows a sample with $\langle q \dot{M} \rangle =1.6\times10^{-11}$, where $q$ is the ion-fraction of Si IV and $\dot{M}$ is the spherically symmetric mass-loss rate of the model \citep[see, e.g.][]{NazSun15}. In preliminary calculations, both the depth of the line and its extent bluewards of line-center depend on the viewing angle, with changes of 3$^\circ$ in viewing angle causing 15-20\% changes in depth of the line and as much as order unity changes in the line width.
However, because fits involve adjustment of several parameters (e.g., for  ionization fraction, macro- and micro-turbulence, viewing angle, etc.), we defer a detailed discussion to a future paper.

\subsection{Model of a rotating B2 star}

\begin{figure}
\centering
\includegraphics[width=0.48\textwidth]{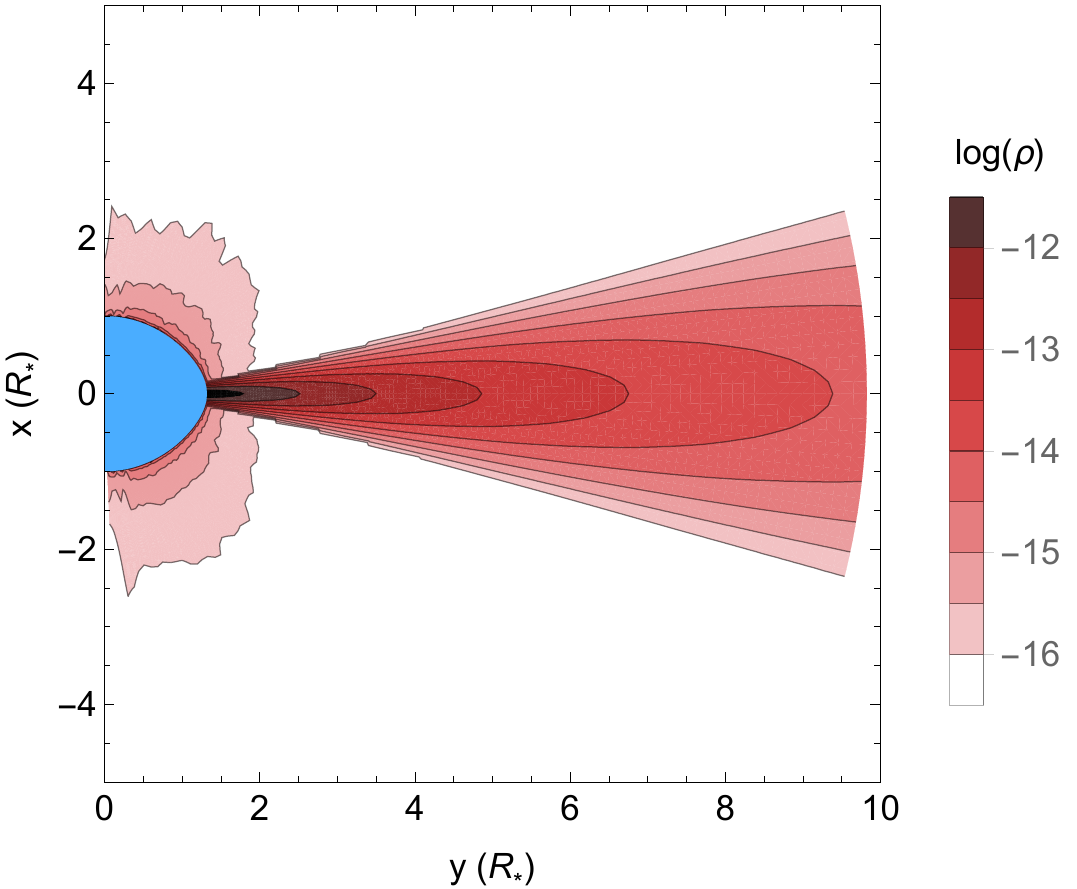}
\caption{
Initial condition of the rotating B2 model, plotted in log density,  in g\,cm$^{-3}$.
}
\label{fig:b2_rot_init}
\end{figure}

Recalling that rapid rotation is a ubiquitous feature of classical Be stars, let us next examine the effect of rapid rotation, and the associated stellar oblateness and gravity darkening, on line-driven ablation.
For this, we modify our standard B2 model to have $W=v_{rot}/v_{orb}=0.8$. Stellar structure dictates that rapid rotation would cause the equatorial radius to swell, leaving the polar radius nearly unchanged. Instead, to isolate the effect of gravity darkening and provide a direct comparison with equatorial disk ablation of a non-rotating model, we here keep the equatorial radius fixed, and so now require the polar radius to shrink.
This means that the rotating and non-rotating models have the same mass and equatorial radius, and thus the same equatorial escape speed.
Figure \ref{fig:b2_rot_init} shows the initial condition superposition of a disk on a relaxed, steady-state, rotating wind. 
Note now the equatorial oblateness of the underlying star, with however a bipolar enhancement of the wind density, reflecting the stronger mass loss from the gravity-brightened poles.

\begin{figure}
\centering
\includegraphics[width=0.48\textwidth]{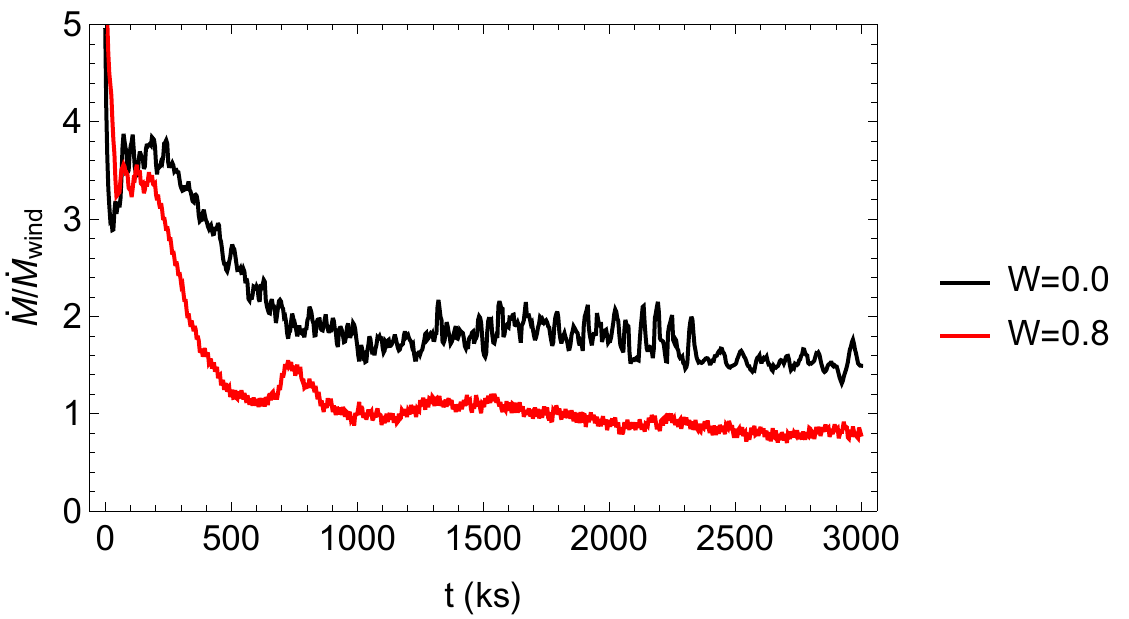}
\caption{
Time evolution of disk ablation rate for both the rotating (W=0.8) and non-rotating (W=0.0) B2 model, in units of the associated steady-state wind mass loss rates (which is about 6\% higher for the non-rotating model).
}
\label{fig:rot_v_norot}
\end{figure}

Figure \ref{fig:rot_v_norot} compares the ablation rate of the models with and without rotation, showing that equatorial gravity darkening modestly reduces the disk ablation rate, by a maximum factor about two. Following the review by \cite{Cra96}, this is consistent with the comparable factor two reduction in equatorial surface brightness between the non-rotating and rotating models, due to this gravity darkening. 
Keeping the equatorial radius constant removes the attendant reduction of equatorial escape speed normally associated with rapid rotation. Therefore, this choice maximizes the difference between the ablation of the rotating and non-rotating models.
In addition, both modern theoretical analyses \citep[see, e.g.][]{EspRie11} and recent interferometric observations \citep{DomKer14} suggest a gravity darkening exponent lower than the value $0.25$ derived by \cite{von24};
adopting such a lower exponent would moderate the effect of gravity darkening on line-driven ablation. 
Taken together, the lower equatorial escape speed and higher equatorial brightness should lead to an ablation rate somewhere between the lower limit presented here and the ablation rate of a non-rotating star.

\subsection{Ablative destruction of a disk around an O7 star}

As discussed above, the incidence of the Be phenomena peaks around spectral type B2 and declines towards earlier spectral types, with only a few detected O9e stars in the Milky Way.
While the exact mechanism for ejection of material into a circumstellar decretion disk is uncertain, we can address here how long such a circumstellar disk can survive the stronger radiative ablation associated with the greater luminosity of such earlier type stars. 
To quantify this, we now present a simulation of the ablation of a marginally optically thin disk around an O7 star, with star, disk, and wind parameters given by tables \ref{tab:o7_params} and \ref{tab:o7_wind} as taken from \cite{MarSch05} and \cite{PulSpr00}. 
The choice of keeping the disk marginally optically thin leads to a factor 5 increase in disk mass, but the factor 25 higher luminosity leads to a factor 200 increase in the wind mass loss rate. Therefore, we should expect much stronger disk ablation.

\begin{table}
\caption{Stellar and Disk Parameters of the O7 Model} \label{tab:o7_params}
\centering
\begin{tabular}{| c | c | c | c | c |}
\hline
 $T_{\rm eff}$ (kK) & $L_\ast$ ($L_\odot$) & $M_\ast$ ($M_\odot$) & $R_\ast$ ($R_\odot$) & $M_{disk}$ ($M_\odot$)\\
\hline
 36 & 1.3$\times 10^5$ & 26.5 & 9.4 & 7.0$\times 10^{-10}$  \\
\hline
\end{tabular}
\end{table}

\begin{table}
\caption{Wind Parameters of the O7 Model} \label{tab:o7_wind}
\centering
\begin{tabular}{| c | c | c | c |}
\hline
 $\bar{Q}$ & $Q_\mathrm{o}$ & $\alpha$ & $\dot{M}_{wind}$ ($M_\odot/yr$) \\
\hline
 2500 & 2200 & 0.66 & 1.5$\times 10^{-7}$ \\
\hline
\end{tabular}
\end{table}

\begin{figure}
\centering
\begin{subfigure}[b]{0.48\textwidth}
\includegraphics[width=\textwidth]{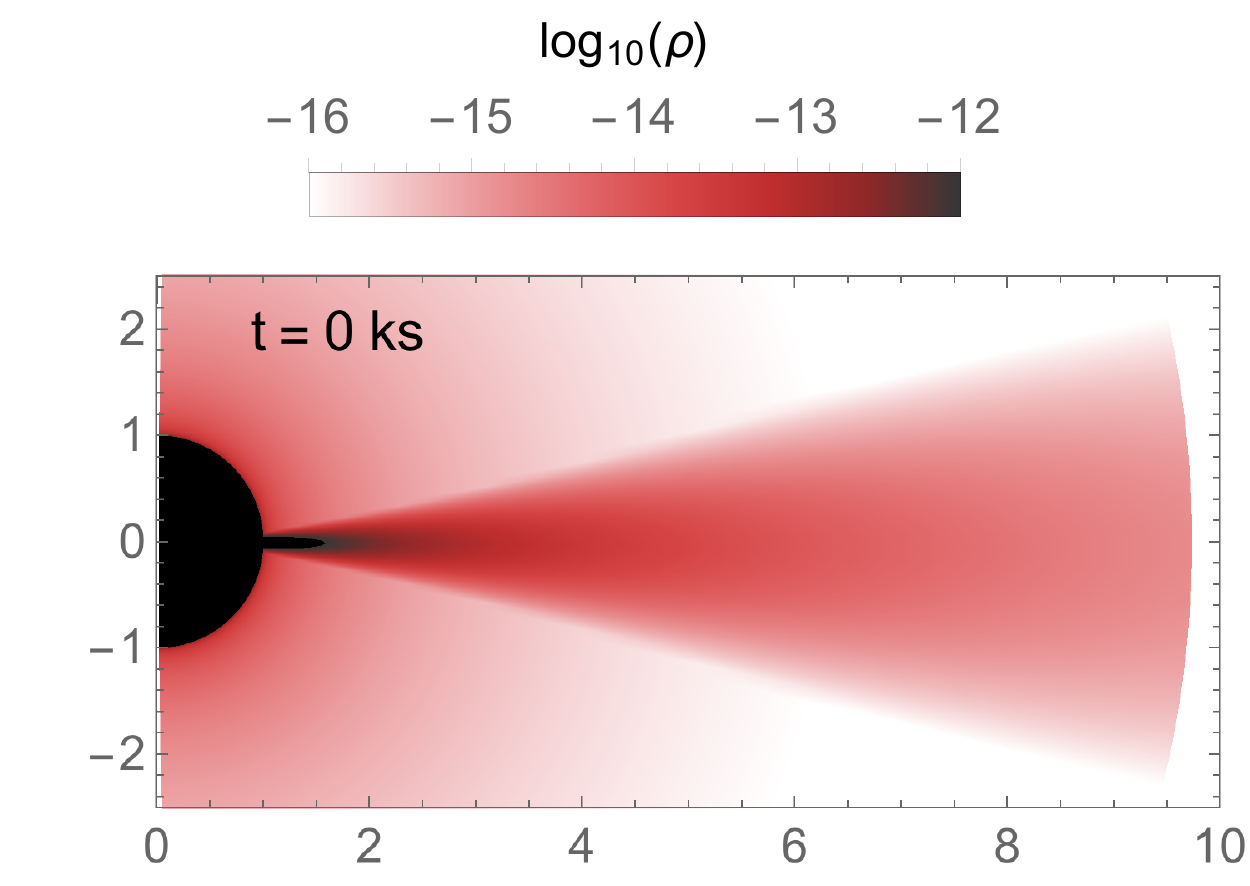}
\end{subfigure}

\begin{subfigure}[b]{0.48\textwidth}
\includegraphics[width=\textwidth]{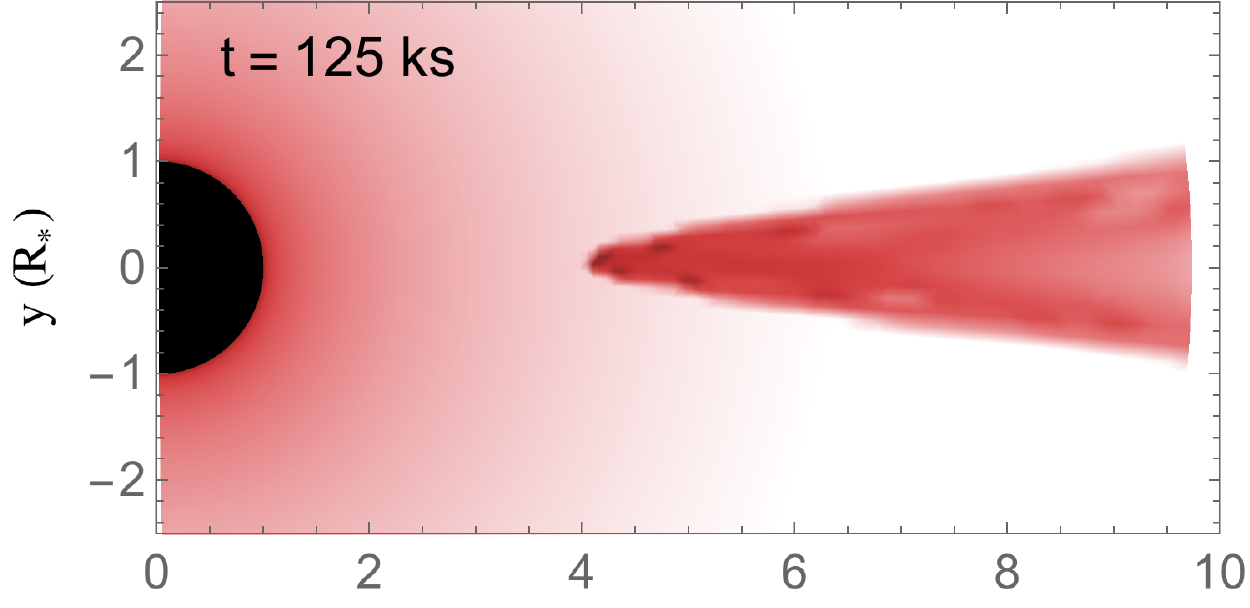}
\end{subfigure}

\begin{subfigure}[b]{0.48\textwidth}
\includegraphics[width=\textwidth]{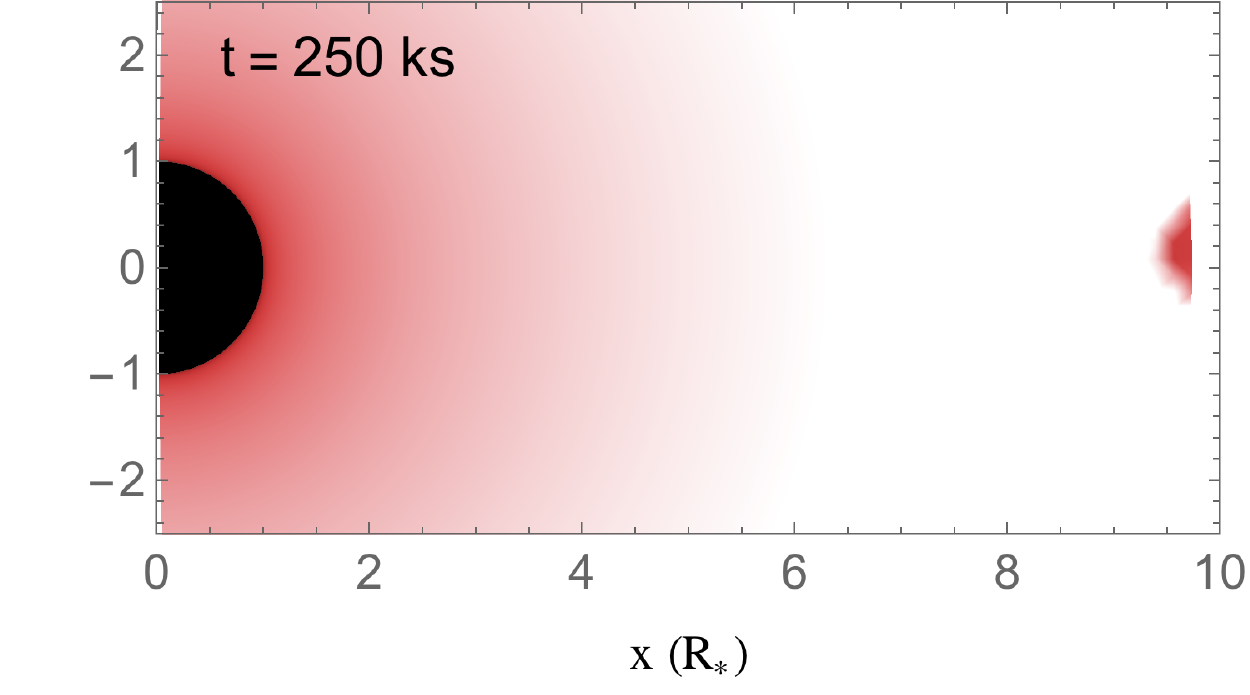}
\end{subfigure}

\caption{
Snapshots of log density for the O7 model at times $t=0$, 125, and 250 ks from the initial condition, showing now the  rapid destruction of the initial disk on a dynamical timescale by the strong radiation field of the O7 star.
}
\label{fig:O7_snaps}
\end{figure}

\begin{figure}
\centering
\includegraphics[width=0.48\textwidth]{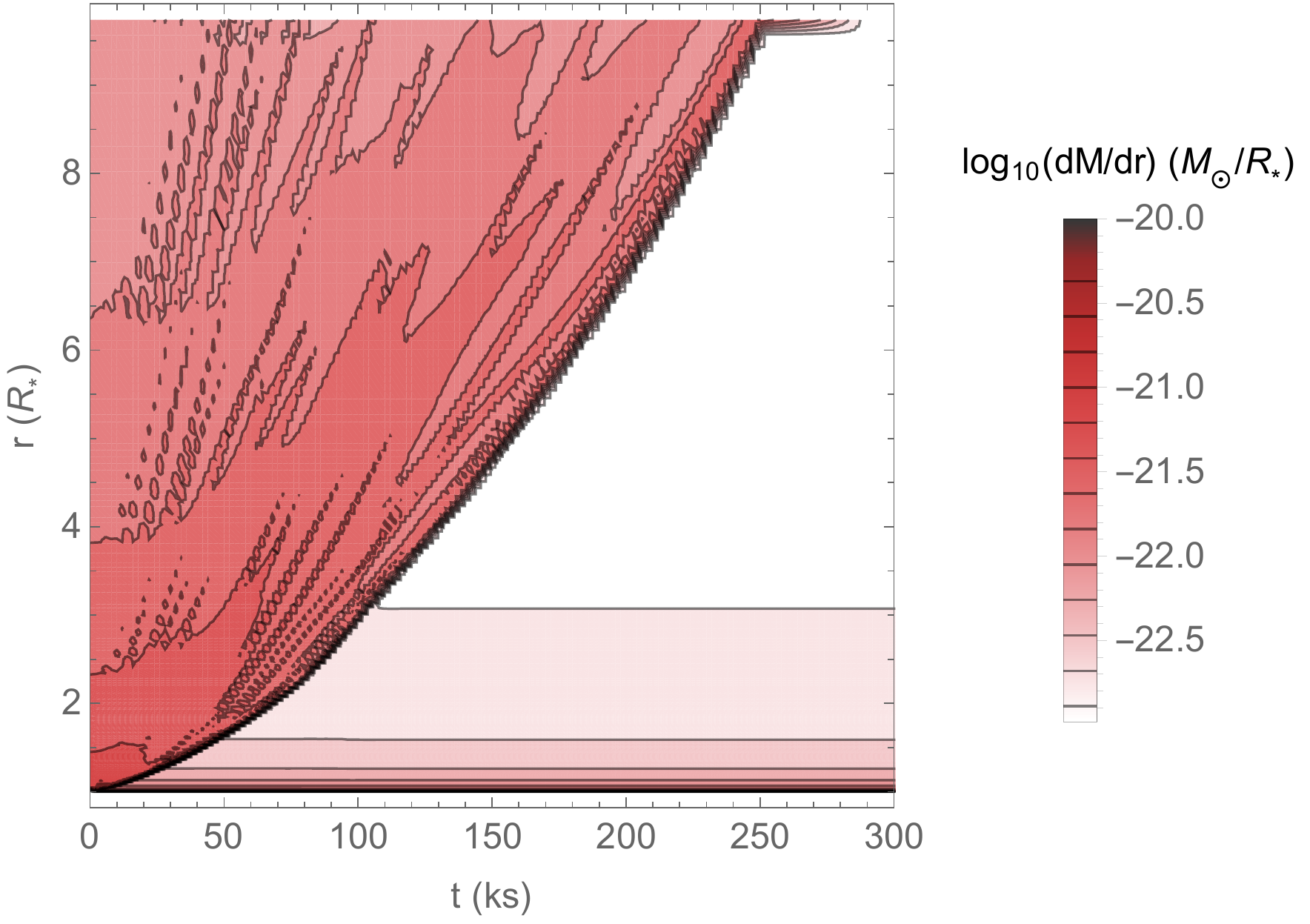}
\caption{
For the O7 model, time evolution of $dM/dr$ (defined in equation \ref{eqn:dMdr}), the mass in spherical shells above the stellar surface, in units of solar masses per stellar radius.
}
\label{fig:O7_dmdr}
\end{figure}

Indeed, instead of the gradual disk surface ablation found for the B2 star case, figure \ref{fig:O7_snaps} shows that  radiative acceleration of an O7 star is strong enough to simply drive away the entire disk over a dynamical timescale. 
To quantify further the evolution of the disk material for this O7-star case,
figure 
\ref{fig:O7_dmdr} plots the spatial and temporal variation of the mass in each spherical shell,
\beq
\frac{dM}{dr}\equiv \oint \rho r^2 d\Omega\,.
\label{eqn:dMdr}
\eeq
Note particularly the rapid removal of material from the inside of the disk outwards, as well as the quite short ($\sim$ three day) time it takes for the disk to be completely evacuated.

The brown curve in figure \ref{fig:mdotdisk} plots the time variation of
the total, wind-normalized ablation rate for this O7 simulation. 
Whereas the B2 model (black curves) slowly relaxes to a relatively steady ablation rate about twice the wind mass loss rate, the O7 model impulsively ejects the disk material in a sustained burst, leaving finally only a steady-state wind. This removal of the disk is so sudden that the ablation rate never significantly exceeds the spherically symmetric mass loss rate.

We thus see that an O7 star destroys a pre-existing optically thin disk on a dynamical timescale. To counter this destruction, the star would need to feed a disk at a rate of over $1.5\times 10^{-7}M_\odot$/yr. This is so much larger than that needed to maintain a disk for later spectral types, and so provides a natural explanation for why there are no galactic O7e stars \citep[see, e.g.][]{MarFre06,RivCar13}.

\subsection{Ablation as a function of spectral type}\label{sec:spec_type}

To bridge between the gradual surface ablation seen for the B2 model and the dynamic disk ejection seen for the O7 star, let us now consider intermediate spectral types, with parameters given in tables
\ref{tab:spec_stars} and \ref{tab:spec_wind}. For the O stars stellar parameters are taken from \cite{MarSch05} while, as was done for the B2 star, \cite{TruDuf07} and \cite{GeoEks13} are used for the B stars. \cite{PulSpr00} is used for all wind parameters.

\begin{table*}
\caption{Stellar and Disk Parameters as a Function of Spectral Type} \label{tab:spec_stars}
\centering
\begin{tabular}{| l | c | c | c | c | c |}
\hline
Sp. type & $T_{\rm eff}$ (kK) & $L_\ast$ ($L_\odot$) & $M_\ast$ ($M_\odot$) & $R_\ast$ ($R_\odot$) & $M_{disk}$ ($M_\odot$) \\
\hline
B3V & 18 & 1.6$\times 10^3$ & 6 & 4.1 & 1.3$\times 10^{-10}$ \\
B2V & 22 & 5.0$\times 10^3$ & 9 & 5.0 & 1.9$\times 10^{-10}$  \\
B1V & 24 & 7.9$\times 10^3$ & 11 & 5.2 & 2.0$\times 10^{-10}$  \\
B0.5V & 28 & 1.6$\times 10^4$ & 13 & 5.5 & 2.3$\times 10^{-10}$  \\
B0V & 31 & 1.6$\times 10^4$ & 16 & 6.0 & 2.7$\times 10^{-10}$  \\
O9V & 32 & 5.0$\times 10^4$ & 18 & 7.7 & 4.8$\times 10^{-10}$  \\
O8V & 33 & 7.9$\times 10^4$ & 22 & 8.5 & 5.7$\times 10^{-10}$  \\
O7V & 36 & 1.3$\times 10^5$ & 26.5 & 9.4 & 7.0$\times 10^{-10}$  \\
\hline
\end{tabular}
\vspace{20pt}
\caption{Wind Parameters as a Function of Spectral Type} \label{tab:spec_wind}
\centering
\begin{tabular}{| l | c | c | c | c |}
\hline
Sp. type & $\bar{Q}$ & $Q_\mathrm{o}$ & $\alpha$ & $\dot{M}_{wind}$ ($M_\odot/yr$) \\
\hline
B3V & 1500 & 7000 & 0.55 & 6.6$\times 10^{-11}$ \\
B2V & 1800 & 4900 & 0.59 & 5.8$\times 10^{-10}$ \\
B1V & 2000 & 4600 & 0.60 & 1.5$\times 10^{-9}$ \\
B0.5V & 2300 & 3900 & 0.63 & 5.7$\times 10^{-9}$ \\
B0V & 2400 & 3400 & 0.64 & 1.9$\times 10^{-8}$ \\
O9V & 2400 & 3300 & 0.65 & 3.6$\times 10^{-8}$ \\
O8V & 2300 & 3100 & 0.65 & 6.9$\times 10^{-8}$ \\
O7V & 2200 & 2500 & 0.66 & 1.5$\times 10^{-7}$ \\
\hline
\end{tabular}
\end{table*}

Given the stellar radius and mass, the base disk density is again set to keep the disk marginally thin, with a fixed radial optical depth $\tau_{disk}=0.4$
Upon integration over the full simulation volume, the scaling of the disk mass is then given by,
\begin{align}
M_{disk}&=\frac{(2\pi)^{3/2}}{\kappa_e} \frac{a}{\sqrt{G M_\ast}} R_\ast^{2.5} \ln(10) \nonumber \\
&=1.24\times10^{-10} M_\odot \sqrt{\frac{T}{10^4 \,\mathrm{K}}}\sqrt{\frac{10M_\odot}{M}}\left(\frac{R}{5R_\odot}\right)^{2.5}  \, .
\end{align}

\begin{figure}
\centering
\includegraphics[width=0.48\textwidth]{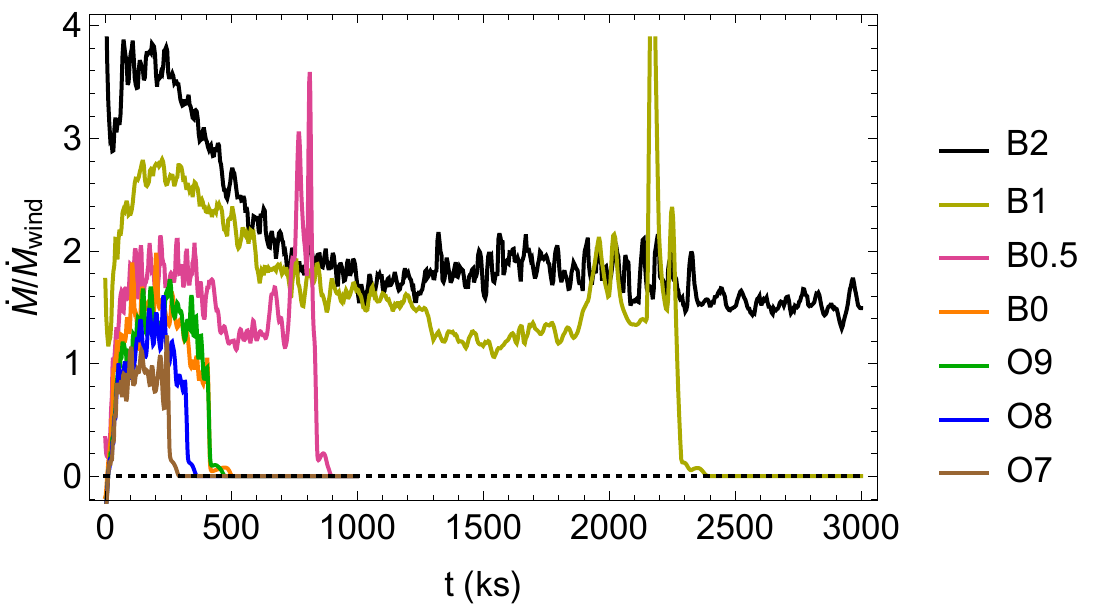}
\caption{
For various spectral types as labeled, the time evolution the disk ablation rate, measured in units of the corresponding spherically symmetric wind mass loss rate $\dot{M}_{wind}$ for each spectral type.
}
\label{fig:mdotdisk}
\end{figure}

For the various spectral types, figure \ref{fig:mdotdisk} compares the time evolution of the wind-scaled disk ablation rate. 
This highlights two populations: 1) Stars B0 and earlier behave like the O7 star, and dynamically eject their disk in a time $t\lesssim 5\times10^5$ s. 2) Stars later than B0 behave like the B2 star, with an interval of steady, slow ablation at a rate comparable to the wind mass loss, terminating in a mass loss burst as the final portion of the disk is removed.

\begin{figure}
\centering
\includegraphics[width=0.48\textwidth]{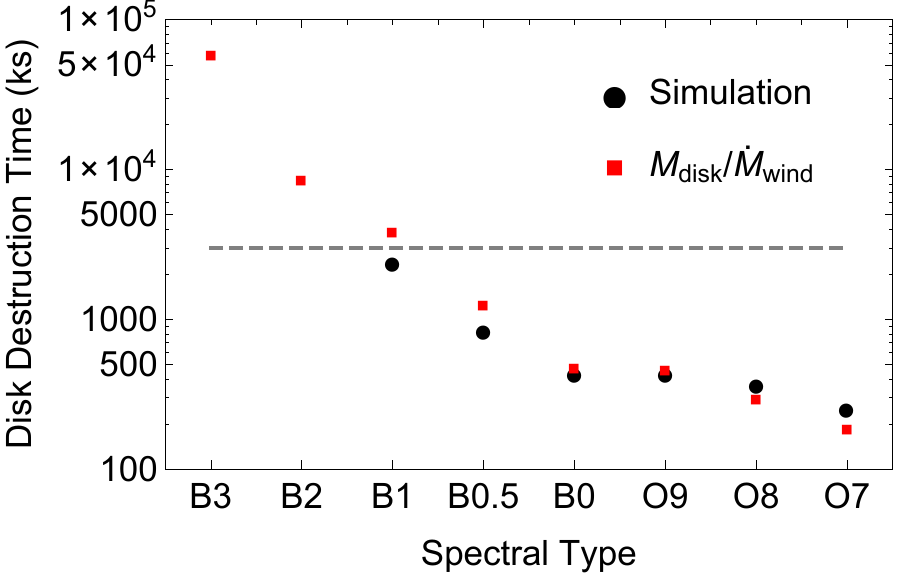}
\caption{
Time to destroy an optically thin disk as predicted by $M_{disk}/\dot{M}_{wind}$ (red squares) compared with the time it actually takes in the simulations (black circles). The dashed line denotes the duration of the longest simulation, meaning no simulations have been run long enough to see the disk removal completed for the B2 or B3 case.
}
\label{fig:t_disk}
\end{figure}

In figure \ref{fig:t_disk}, the black circles plot the disk destruction time for each of the various spectral types. These follow closely the simple scaling given by the ratio of the initial disk mass to wind mass loss rate $M_{disk}/\dot{M}_{wind}$, shown in red squares. This suggests that this simple scaling form can be generally used to predict disk destruction times, even for later spectral types and longer simulation times than considered here. Since the disk mass diverges only logarithmically with outer disk radius, increasing the simulation volume will increase the disk mass, and thus the disk destruction time, by an order unity factor at most. However, as most Classical Be stars have emitting regions inferred to be on the order of, or less than, 10$R_\ast$ \citep[see, e.g.][]{MeiMil12, TouGie13}, and since disk removal occurs from the inner edge outwards, the time scales presented here will still provide a good estimate of the time for circumstellar emission to disappear.

The ability of this simple scaling to predict simulation results guides its use to predict the behavior of actual Be stars.
Previous work by \cite{CarBjo12} observationally inferred a disk decay for the Be star 28 CMa of order a half year. To reproduce this disk decay time within the standard viscous diffusion model, these authors invoked a viscosity coefficient\footnote{Reanalysis of the data suggests a potential modest reduction to $\alpha\approx0.4$. (Carciofi, private communication)} $\alpha\approx1$ \citep{ShaSun76}. By comparison, simulations of the magneto-rotational instability, thought to be the origin of viscous transport in ionized gaseous disks, show a much lower value, $\alpha \lesssim 0.01$ \citep[see, e.g.][]{PapNel03}, implying then a much longer diffusive decay time. The inferred stellar parameters of 28 CMa from \cite{MaiRiv03} are very similar to the B2 model, for which the predicted ablative disk destruction time is within a factor of two of the observationally inferred disk decay time.

This important result suggests that line-driven ablation can readily explain the relatively short, few-month timescale observationally inferred for Be-disk decay, without the need to invoke anomalously strong viscous diffusion.

\section{Summary and Future Work}
\label{sec:conclusions}

The study here has carried out detailed multi-dimensional radiation-hydrodynamical simulations of line-driven ablation of optically thin disks around luminous, early-type stars. A key general result is that the disk mass ablation rate scales closely, within a factor of a few, with the global mass loss rate of the steady-state stellar wind, implying that the disk destruction time scale can be characterized by the simple ratio of disk mass to steady-state wind mass loss rate, $M_{disk}/\dot{M}_{wind}$. 
For O stars, this implies disk destruction on a dynamical timescale, thus helping to explain the relative rarity of classical Oe stars. For Be stars, the destruction takes much longer than the dynamical timescale, but is still much shorter than a characteristic viscous diffusion time, thus helping to explain the inferred rapid decay of Be disks without having to appeal to anomalously strong viscous diffusion.

In the context of Be stars, future work should apply the results and scalings derived here to other cases where there is disk decay after a presumed cessation of the processes that feed the disk. More generally, however, future work should seek to identify mechanisms for disk feeding, perhaps through mass ejection induced by stellar pulsations \citep[see][]{KeeOwo14b}, and couple these with simulations of disk decay through radiative ablation.
As discussed in section \ref{sec:shelluvcomp}, a detailed comparison of synthetic and observed UV line profiles from Be shell stars can provide direct diagnostic constraints on the disk surface ablation models computed here.

Future work can also extend these results to lower metallicity environments.
The Magellanic clouds show a higher fraction of Be stars \citep[e.g.,][]{MarFre06,MarFre07}, 
with the Classical Be phenonemon also extending to much earlier spectral types than in the Milky Way \citep{GolOey16}.
In the context of line-driven disk ablation, these differences could be a natural consequence of the lower metallicity and attendant lower effective opacity in UV metal ion lines. It is well-established both observationally and theoretically \citep[e.g.,][]{VindeK01, MokdeK07} that line-driven winds are weaker at lower metallicites, 
and, given the results presented here, we can expect weaker disk ablation in the lower-metallicity Magellanic clouds.

In a broader context of accretion disks from pre-main sequence stars, a key need is to generalize the above models to account for the effects of significant continuum absorption and scattering in their denser, optically thick disks,
where cooling to temperatures significantly below $T_{\rm eff}$ will also occur.
This will allow exploration of the role of line-scattering accelerations in mass accretion and massive star formation. These results could even have potential implications for controlling the stellar upper mass limit.

\section*{Acknowledgements}
We acknowledge use of the {\tt VH-1} hydrodynamics code, developed by J.~Blondin and collaborators.
This work was supported in part by NASA ATP grant NNX11AC40G and  NASA Chandra grant TM3-14001A, awarded to the University of Delaware.
J.O.S. acknowledges funding from the European Union's Horizon 2020 research and innovation programme under the Marie Sklodowska-Curie grant agreement No 656725.
Finally, we acknowledge many helpful comments and constructive criticisms from the anonymous referee.

\appendix
\section{Multiple Line Resonances}
\label{sec:appa}

\subsection{Analytic wind+disk velocity model}

For our models of a fast stellar wind at high latitudes and a equatorial Keplerian disk with reduced or no radial expansion, lines of sight from the stellar surface toward 
and/or through 
the disk can have a non-monotonic line-of-sight velocity, potentially leading then to multiple Sobolev line resonances with equal line-of-sight speed.
To characterize the potential effect of shadowing by a more inward resonance on the radiative driving within or near the surface of the equatorial disk, let us consider here a simplified analytic form for such a disk+wind model.

\begin{figure}
\includegraphics[width=0.48\textwidth]{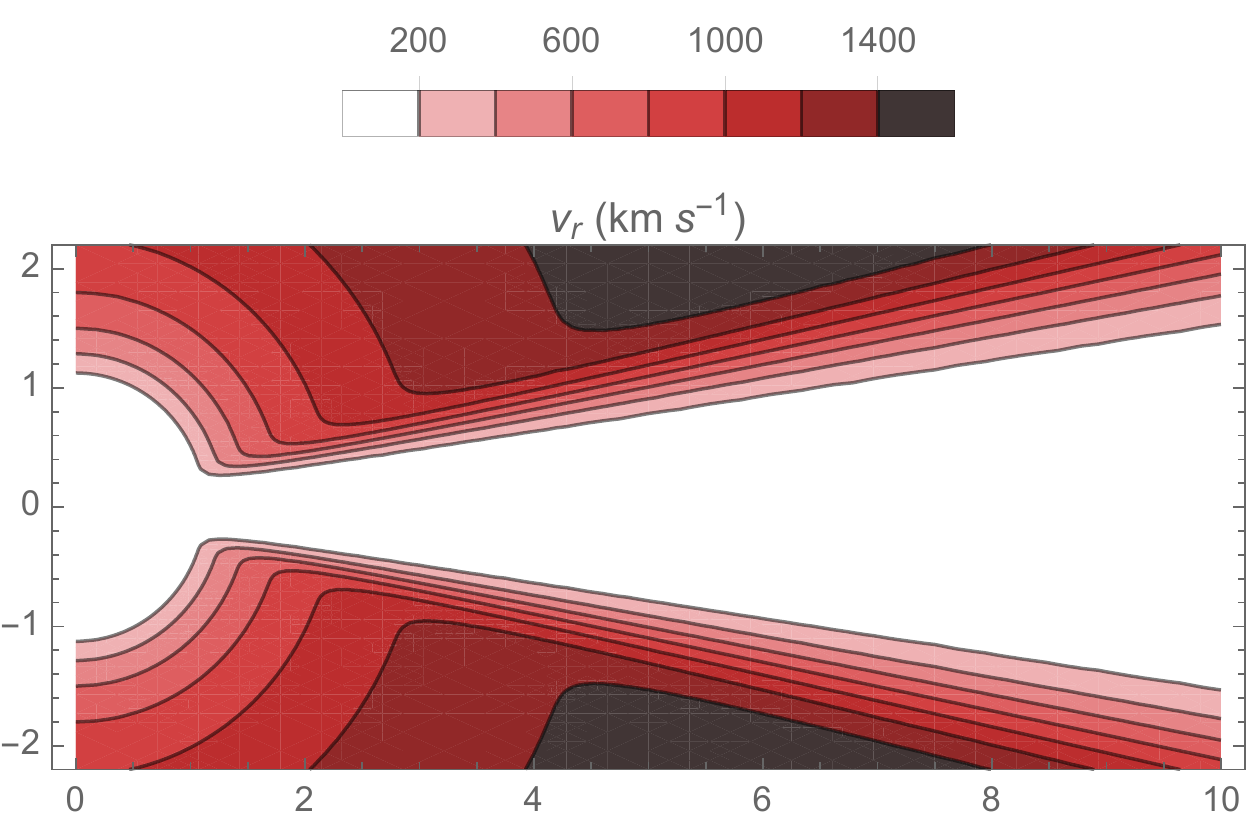}
\caption{
Contours of the spatial variation of radial velocity $v_r$ for analytic wind+disk model with parameters $\theta_d =12^\circ$, $\Delta \theta = 4^\circ$ and $V_{\infty,o} = 1800$\,km\,s$^{-1}$, 
chosen to correspond roughly to the velocity variation in the simulation snapshot shown in figure \ref{fig:B2_vel_snap}.
}
\label{fig:vr-anal}
\end{figure}

Specifically, in terms of
the radius $r$ and 
latitude\footnote{To center on the equatorial disk, we here take $\theta$  to be the {\em latitude}, instead of the {\em co}-latitude used in standard spherical coordinates.}
$\theta$,
 let us assume the radial component the flow speed $v_r$ follows a simple variation\footnote{Often dubbed a ``beta velocity law", with here the velocity exponent $\beta=1$.} from the surface radius $\Rstar$,
\beq
v_r (r,\theta ) =  V_{\infty} (\theta) \left ( 1 - \frac{\Rstar}{r} \right ) \, ,
\label{eq:vrtheta}
\eeq
with however a terminal speed $\vinf (\theta )$ that has a fixed high value $V_{\infty,o}$ over the poles, but then declines steeply near the disk, approaching zero at the equatorial midplane.
Specifically, for a disk-wind boundary of half-width $\Delta \theta$ centered on latitudes $\pm \theta_d$, we take
\beq
V_{\infty} (\theta) = V_{\infty,o} \, \half {\rm Erfc} ((\theta_d - |\theta | )/\Delta \theta )
\, .
\label{eq:vitheta}
\eeq
To approximate the velocity form given in figure \ref{fig:B2_vel_snap} for an actual simulation snapshot, we take here $\theta_d =12^\circ$, $\Delta \theta = 4^\circ$ and $V_{\infty,o} = 1800$\,km\,s$^{-1}$;
figure \ref{fig:vr-anal} then shows that the radial velocity in this analytic model has a similar  form to that of the numerical simulation snapshot in figure \ref{fig:B2_vel_snap}.

The disk component is taken to have an azimuthal speed given by the  Keplerian form for equatorial orbit, 
\beq
v_\phi (r,\theta ) = V_{\rm orb} \sqrt { \frac{\Rstar}{r \cos \theta  } }
\, ,
\label{eq:vphi}
\eeq
where $V_{\rm orb} = \sqrt{GM/\Rstar}$ is the orbital speed at the equatorial surface radius $\Rstar$, taken here to be a third of the polar wind speed, i.e. $V_{\rm orb} = 600$\, km\,s$^{-1}$.

\begin{figure}
\begin{center}
\includegraphics[width=0.36\textwidth]{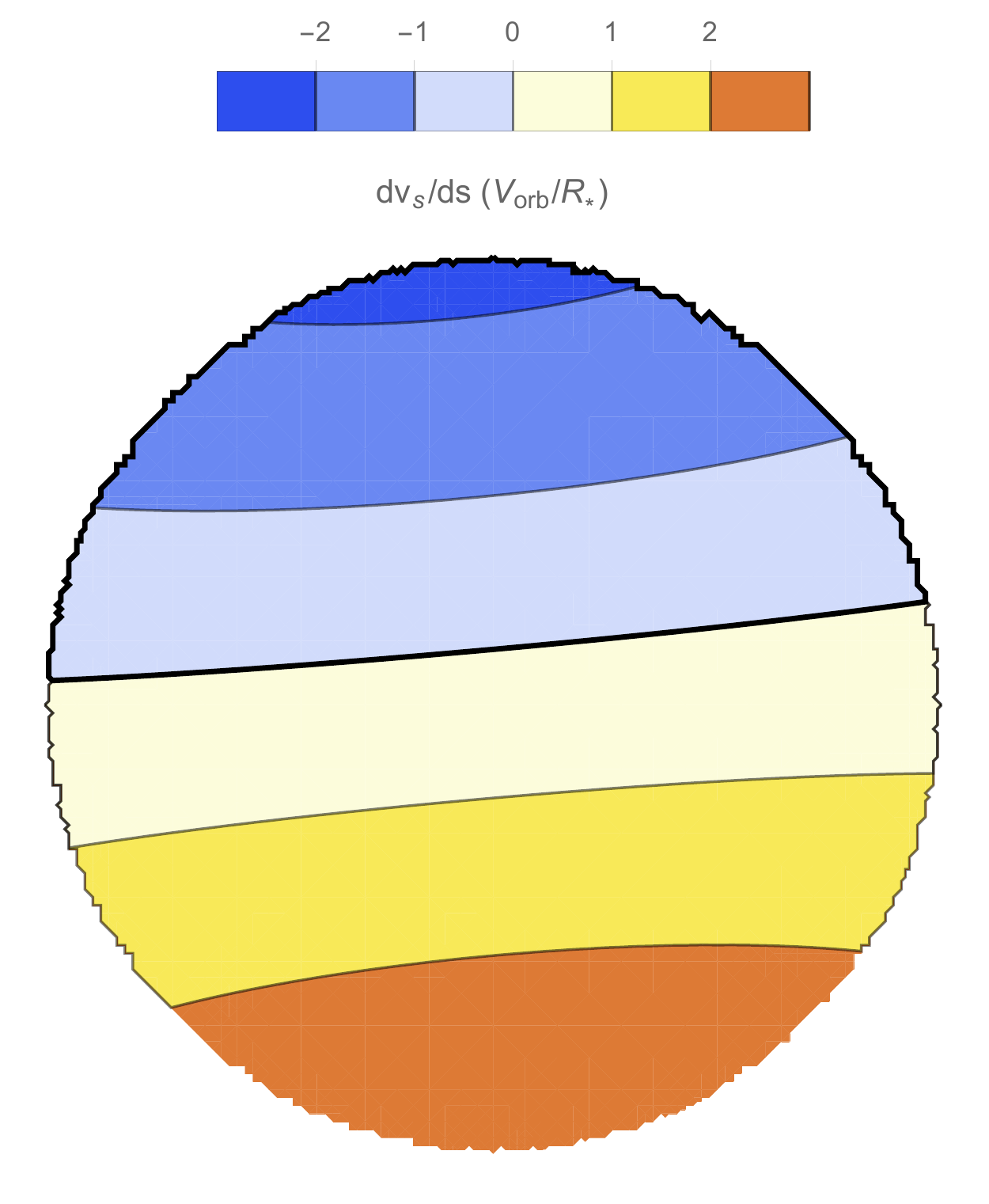}
\end{center}
\vskip -0.2in
\caption{
Contour map on the stellar core of projected velocity gradient $dv_s/ds$ (in units of $V_{\infty,o}/\Rstar$) along lines of sight ${\hat s}$ from the star to a radius $r_p=2 \Rstar$ and latitude $\theta_p=10^\circ$, a position in the middle of the wind-disk boundary layer.
}
\label{fig:dvsds}
\end{figure}

\begin{figure*}
\includegraphics[width=0.3\textwidth]{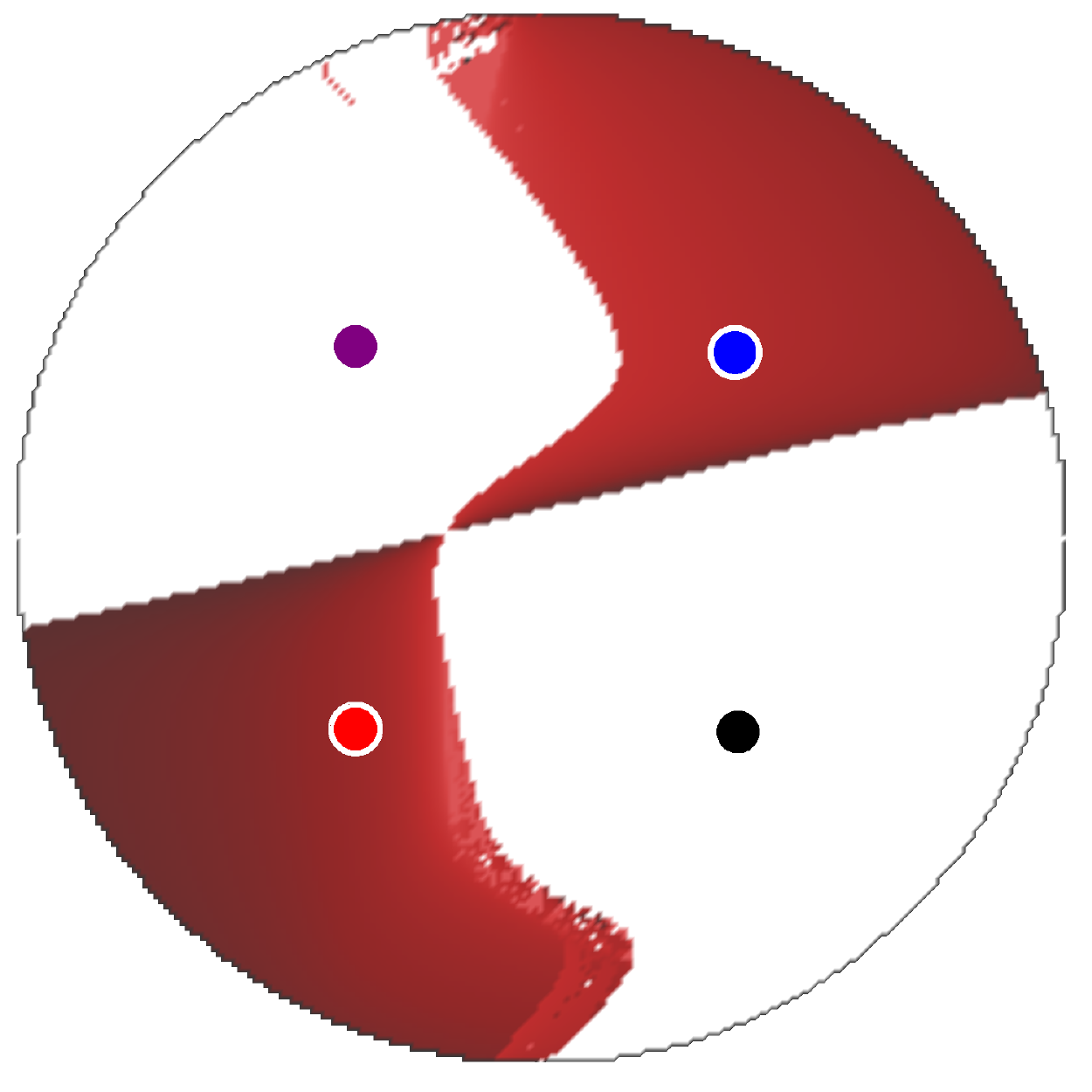}
\includegraphics[width=0.3\textwidth]{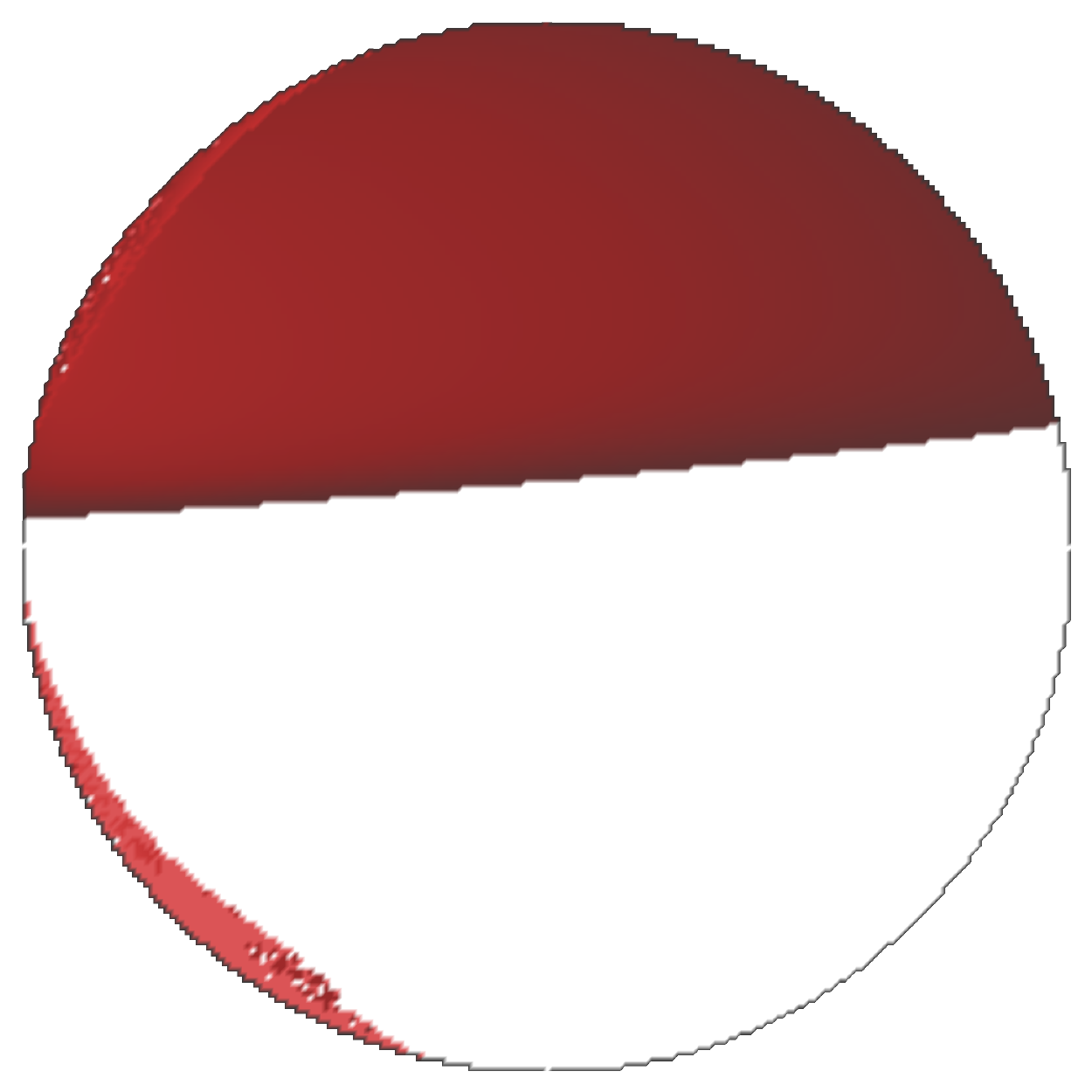}
\includegraphics[width=0.3\textwidth]{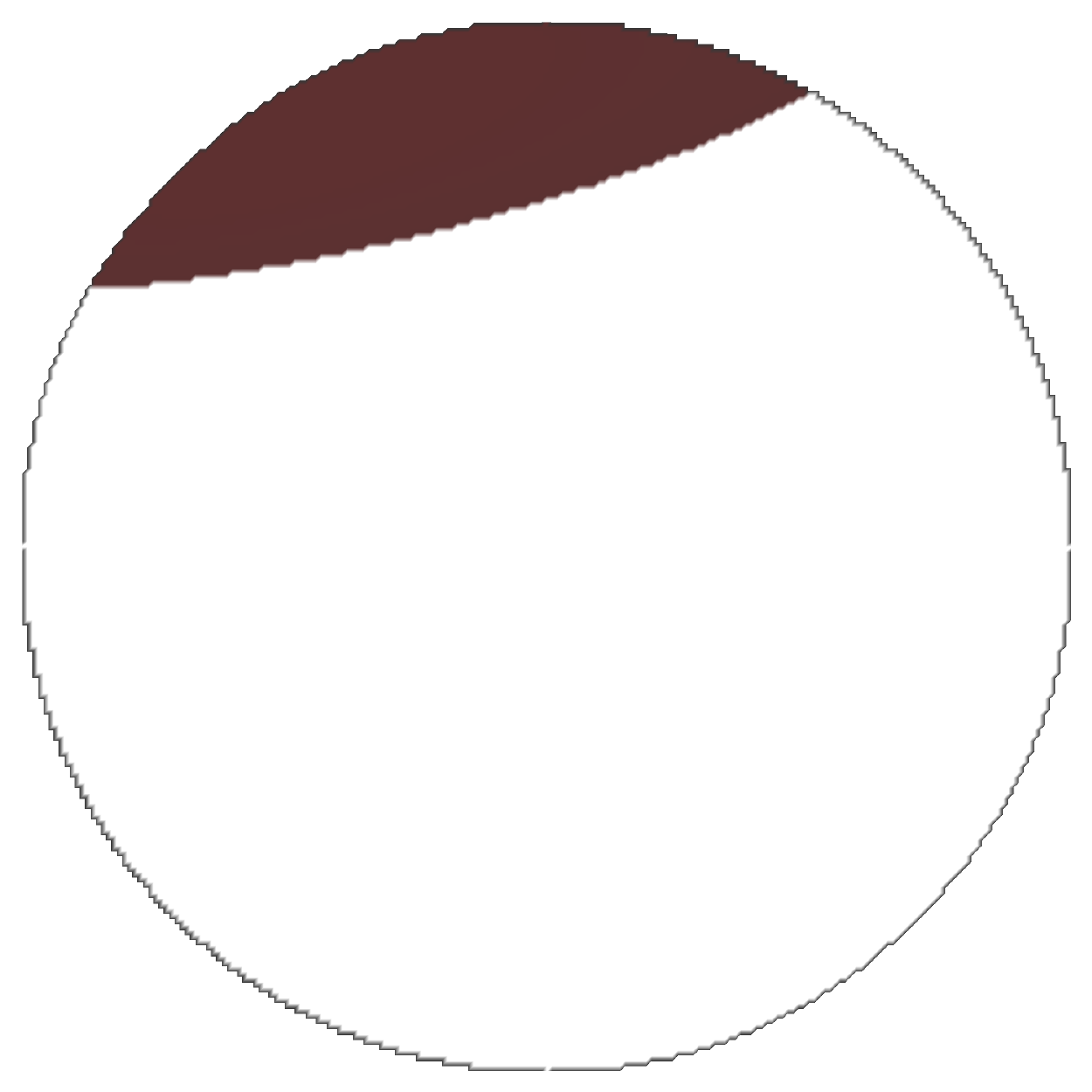}
\includegraphics[width=0.04\textwidth]{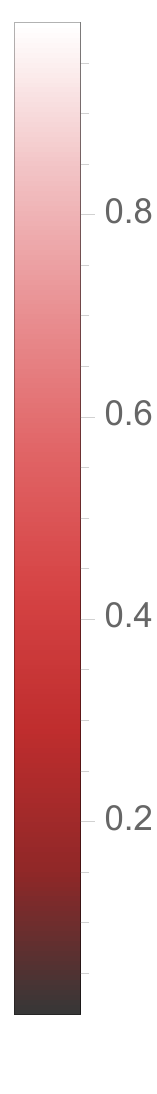}
\caption{
Color maps on the stellar core of intensity $I_o$ (in units of core intensity $I_{core}$) illuminating the positions at radius $r_p=2 \Rstar$, with the left, center and right panels showing results for position latitudes $\theta_p=8^\circ$, $10^\circ$, and $12^\circ$  that are below,  within, and above the wind-disk boundary layer.
The four colored dots in the leftmost panel represent sight lines for the projected velocity plots in figure \ref{fig:vsphip}.
}
\label{fig:Io}
\end{figure*}

\begin{figure}
\includegraphics[width=0.48\textwidth]{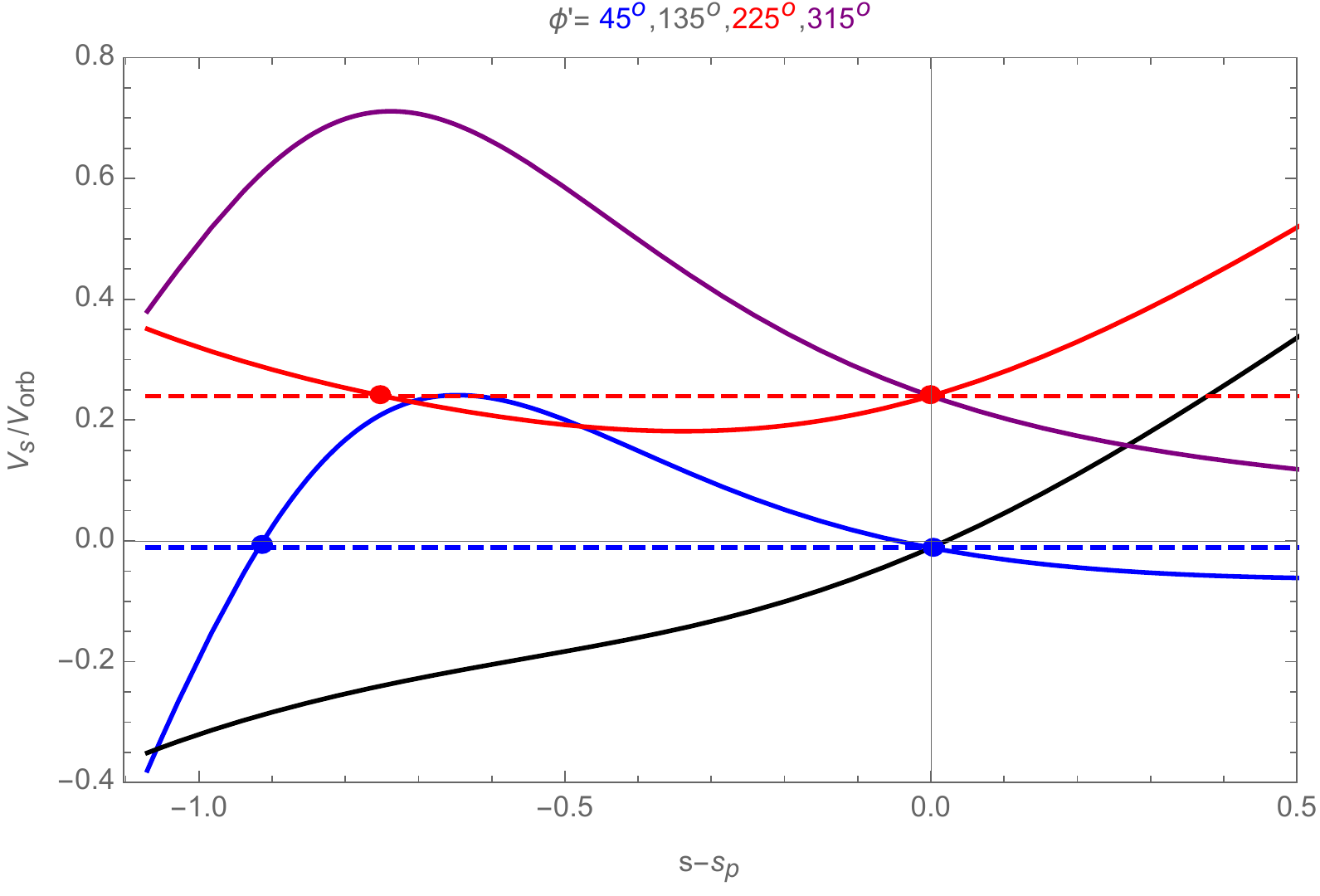}
\caption{
For that same disk viewpoint ($r_p=2 \Rstar$;  $\theta_p = 8^\circ$) used for the leftmost panel of figure \ref{fig:Io},  line plots of the spatial variation of velocity component $v_s$ along the four sight lines toward the stellar core shown by the colored dots in that figure.
These all have impact parameter $p=0.5 \Rstar$ from stellar disk center, but with distinct 
orientations set by polar angles $\phi'=$ 45$^\circ$, 135$^\circ$, 225$^\circ$, and 315$^\circ$ 
(solid lines colored blue, black, red, and purple), representing stellar quadrants ranging clockwise 
from upper right to upper left.
The red and blue dashed curves connect the local velocity at the view point to a common projected velocity at an inner resonance closer to the star, as signified by the red and blue dots.
The black and purple curves have no such inner resonance.
}
\label{fig:vsphip}
\end{figure}

\subsection{Projected velocity gradient and shadowing by an inner resonance}

Let us now consider conditions at some fixed wind+disk position with radius $r_p $ and co-latitude $\theta_p$.
For a near-star position $r_p=2 \Rstar$ in the wind-disk boundary layer with $\theta_p = \theta_d = 12^\circ$, figure \ref{fig:dvsds} plots a contour map on the stellar disk of the line-of-sight velocity gradient 
$dv_s/ds \equiv ({\hat s} \cdot \nabla) ({\bf v} \cdot {\hat s})$ for directional unit vectors ${\hat s}$ from the stellar surface to the position $\{ r_p, \theta_p \}$.
The heavy black contour for $dv_s/ds=0$ encircles the (blue-shaded) northern region with negative gradient, $dv_s/ds <0$, 
since these rays cross from the fast wind at high-latitude to the slower disk-boundary layer.
This is opposite to the strictly positive gradients found in a spherically symmetric, expanding stellar wind, and the associated non-monotonic velocity raises the possibility of multiple line resonances along these rays.
The shadowing of the stellar radiation by an inner resonance can reduce the line-driving in the disk and/or its surface layers.
For optically thick lines, the shadowed intensity illuminating the wind-disk point is set by the source function at the line resonance,
$I_o (r_p,\theta_p) = S ({\bf r}_{res})$.

To quantify this effect, we next use standard root finding methods to locate such inner resonances ${\bf r}_{res}$. In principle, determining
the resonant source function $S ({\bf r}_{res})$ requires a self-consistent treatment of all the nonlocal couplings among global multiple resonances. Instead, we choose here to simply characterize this in terms of an {\em optically thin} source function that merely accounts for the geometrical dilution of the mean intensity from the stellar core, giving then
\beq
I_o (r_p,\theta_p) \approx S_{thin} (r_{res}) \equiv I_{core} \left [ 1 - \sqrt{1 - (\Rstar/r_{res})^2 } \right ]
\, .
\label{eq:Io}
\eeq
Unshadowed rays without any such inner resonance still have $I_o = I_{core}$; ignoring any limb darkening, we  take this to be constant over the stellar disk.

For this same wind-disk boundary point ($r_p=2 \Rstar$, $\theta_p=12^\circ$) taken for figure \ref{fig:dvsds} , the central panel of figure \ref{fig:Io} maps contours of the illuminating intensity $I_o$ from the associated directions toward the stellar disk. All  the northern hemisphere regions with a negative line-of-sight gradient in figure \ref{fig:dvsds} have reduced intensity, due to the shadowing by the inner resonance. But note that there is also a sliver along the left limb in the south where the intensity is also reduced, indicating that even these directions with a positive velocity gradient at the  specified wind-disk point have an interior resonance that reduces the intensity. 

Nonetheless, over more half the stellar disk, mostly in the southern hemisphere that illuminates the point from below, through the disk,  there is no inner resonance. The unattenuated illumination from these regions thus can sustain the line-driving, albeit with a reduction in strength. In this example, this amounts to a flux reduction factor of 0.65 if one assumes the thin source function at the resonance, or a factor 0.54 if one assumes complete absorption at this resonance.

As shown in the right panel of figure \ref{fig:dvsds}, at the higher latitude $\theta_p = 16^\circ$ above the disk-wind boundary, the covering fraction in the northern hemisphere shrinks,  weakening the line driving reduction.
Indeed,  for $| \theta_p | > 18^\circ$, one recovers full illumination without any attenuation.

At lower latitudes within the wind-disk boundary, the Keplerian shear of the disk plays an important role in the formation or inhibition of an inner resonance. The left panel of figure \ref{fig:Io} shows that for latitude $\theta_p = 8^\circ$ this leads to an increase in the shadowing arc on the left side of the southern hemisphere, but with a corresponding {\em decrease} in the resonance covering in the left side of the northern hemisphere.

For this disk viewpoint, figure \ref{fig:vsphip} plots the spatial variation of light-of-sight velocity for sight lines toward the four quadrants of the stellar disk.
This illustrates how the sight lines toward the upper right (blue) and lower left (red) quadrants have an inner resonance (connected by dashed lines), while the two other sight lines have no such resonance, even though the velocity along the sight to the upper right quadrant is non-monotonic, with a negative local velocity gradient at the considered viewpoint.

\subsection{Flux reduction}

Finally, by integrating the illumination intensity $I_o$ over the area of the stellar disk for any range of wind/disk viewpoints, we can compute the spatial dependence of the associated flux reduction factor.
This then provides an estimate of the associated reduction of the local radiative driving relative to a computation that, as in our simulation models, neglects the effect of multiple line resonances.

\begin{figure}
\includegraphics[width=0.47\textwidth]{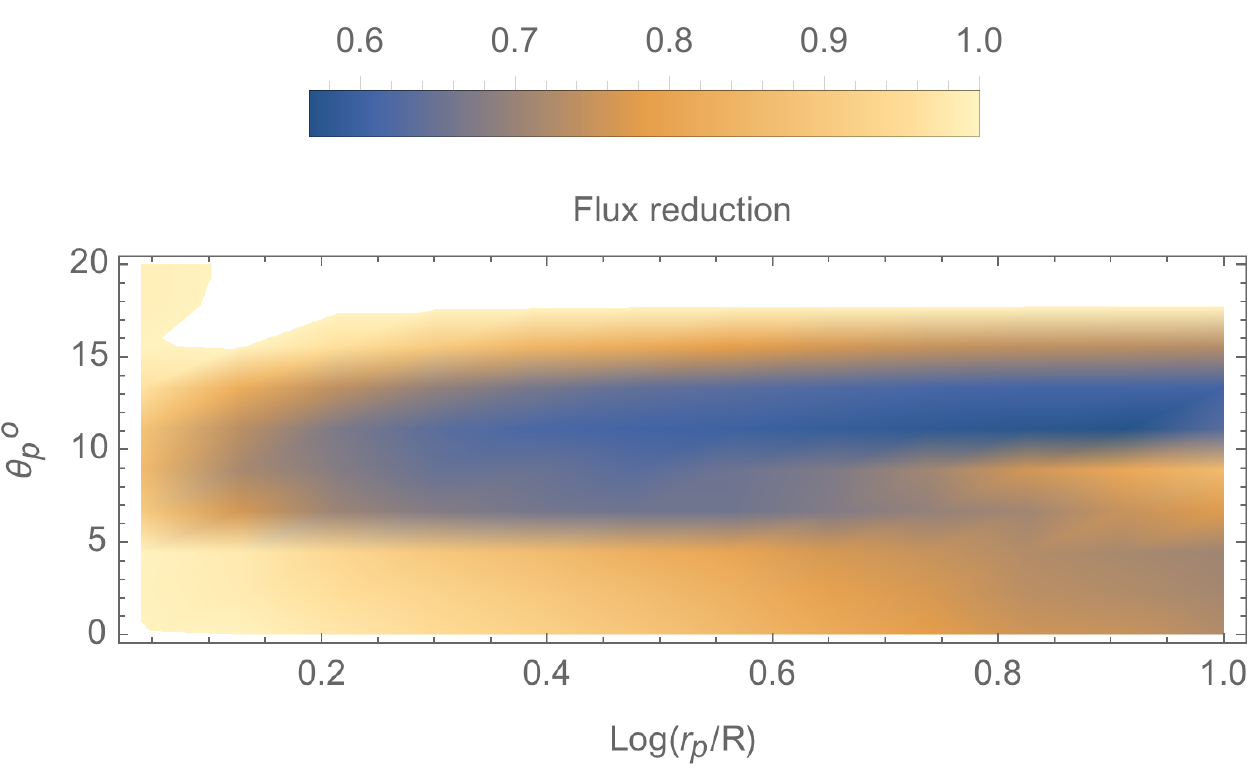}
\caption{
Color map of the flux reduction from inner resonance shadowing, plotted vs. log radius $\log ( r_p/\Rstar ) $ 
over latitudes  from the equator to a maximum $ \theta_p = 20^\circ > \theta_d$ above the wind-disk boundary.
}
\label{fig:flux}
\end{figure}

Figure \ref{fig:flux} shows a color map of this flux reduction factor, plotted as a function of wind/disk point radius (on a log scale) and latitude.  
Note that the strongest reduction, with a value just below 0.6, occurs along a band centered on the wind-disk boundary latitude $\theta_d = 12^\circ$. 
The reduction weakens near the stellar surface, and toward the equator, and approaches unit value at high latitudes $\theta_p > 18^\circ$ above the wind-disk boundary.

If one assumes optically thick pure {\em absorption} at the inner resonance -- instead of  an optically thin source function for forward-scattered radiation --, then the flux reduction is generally about 10\% stronger.
On the other hand, a solution that self-consistently accounts for the scattered radiation coupling among global resonances could increase this inner resonance source function, with then a weakening of the associated flux reduction in radiative driving.

The overall upshot is then that accounting for multiple resonance in the wind+disk simulations here should result  in at most a modest, factor-two reduction in the line force, with the greatest effect concentrated around wind-disk boundary layer.
The reduction should be weaker in the disk midplane, and approach unity above the disk in the high-speed wind.

\bibliographystyle{mn2e}
\bibliography{biblio}

\end{document}